\documentclass[10pt,a4paper]{amsart}

\usepackage{amssymb,eucal,mathrsfs,array,setspace,geometry,enumitem,cite}
\usepackage{thmtools,thm-restate} 
\usepackage{mathtools}
\usepackage[dvipsnames]{xcolor}
\usepackage{graphicx}
\usepackage{tikz}
\usetikzlibrary{decorations.pathreplacing}
\usepackage[colorlinks=true,citecolor=black,linkcolor=black,urlcolor=black]{hyperref}

\setitemize{leftmargin=*}   
\setenumerate{leftmargin=*, label=\textup{(\arabic*)}} 

\let\originalleft\left     
\let\originalright\right
\renewcommand{\left}{\mathopen{}\mathclose\bgroup\originalleft}
\renewcommand{\right}{\aftergroup\egroup\originalright}

\newcolumntype{C}{>{$}c<{$}} 

\numberwithin{equation}{section}
\allowdisplaybreaks

\def\be{\begin{equation}}
\def\ee{\end{equation}}

\DeclareMathOperator{\tr}{tr}

\newcommand{\dd}{\mathrm{d}}   
\newcommand{\ii}{\mathfrak{i}} 

\newcommand{\alg}[1]{\mathfrak{#1}}  
\newcommand{\grp}[1]{\mathsf{#1}}    
\newcommand{\fld}[1]{\mathbb{#1}}

\newcommand{\ZZ}{\fld{Z}}

\newcommand{\RR}{\fld{R}}
\newcommand{\CC}{\fld{C}}

\DeclarePairedDelimiter{\normord}{{} :}{: {}} 
\newcommand{\Or}{{\mathcal O}}   

\def\T{\mathcal T}  
\def\tT{\tilde{\mathcal T}}  

\def\ta{\theta}
\def\Han{H_{\mathrm{an}}}  
\def\Qan{Q_{\mathrm{an}}}  
\def\Pian{\Pi_{\mathrm{an}}^{\phantom \dagger}} 
\def\Piand{\Pi_{\mathrm{an}}^{\dagger}} 
\def\phian{\phi_{\mathrm{an}}^{\phantom \dagger} } 
\def\phiand{\phi_{\mathrm{an}}^{\dagger} } 
\def\ep{\epsilon}
\def\h#1{\hat{#1}}
\def\lt#1{\left#1}
\def\rt#1{\right#1}
\def\p{\partial}

\def\bphi{\phi^\dagger}

\def\FTG{\mathscr{T}} 

\def\TG{\mathsf T} 

\begin{document}

\title[]{Expectation values of twist fields and universal entanglement saturation of the free massive boson}

\author[O Blondeau-Fournier]{Olivier Blondeau-Fournier}

\address[Olivier Blondeau-Fournier]{
Department of Mathematics \\
King's College London \\
Strand, United Kingdom, WC2R~2LS.
}

\email{olivier.blondeau-fournier@kcl.ac.uk}

\author[B Doyon]{Benjamin Doyon}

\address[Benjamin Doyon]{
Department of Mathematics \\
King's College London \\
Strand, United Kingdom, WC2R~2LS.
}

\email{benjamin.doyon@kcl.ac.uk}

\thanks{\today}

\dedicatory{Dedicated to the memory of Petr Petrovich Kulish}

\begin{abstract}
The evaluation of vacuum expectation values (VEVs) in massive integrable quantum field theory (QFT) is a nontrivial renormalization-group ``connection problem" -- relating large and short distance asymptotics -- and is in general unsolved. This is particularly relevant in the context of entanglement entropy, where VEVs of branch-point twist fields give universal saturation predictions. We propose a new method to compute VEVs of twist fields associated to continuous symmetries in QFT. The method is based on a differential equation in the continuous symmetry parameter, and gives VEVs as infinite form-factor series which truncate at two-particle level in free QFT. We verify the method by studying $\grp{U}(1)$ twist fields in free models, which are simply related to the branch-point twist fields. We provide the first exact formulae for the VEVs of such fields in the massive uncompactified  free boson model, checking against an independent calculation based on angular quantization. We show that logarithmic terms, overlooked in the original work of Callan \& Wilczek [Phys.~Lett.~B333 (1994)], appear both in the massless and in the massive situations. This implies that, in agreement with numerical form-factor observations by  Bianchini \& Castro-Alvaredo [Nucl.~Phys.~B913 (2016)], the standard power-law short-distance behavior is corrected by a logarithmic factor. We discuss how this gives universal formulae for the saturation of entanglement entropy of a single interval in near-critical harmonic chains, including $\log\log$ corrections.
\end{abstract}

\maketitle

\onehalfspacing

\section{Introduction} 

The solution to a quantum field theory (QFT) model can be understood as the full set of its local fields and their correlation functions. In massive QFT, vacuum correlation functions may be evaluated in two complementary ways: using the spectral expansion, giving rise to fast-converging large-distance expansions; or using conformal field theory (CFT) and perturbations thereof, providing systematic short-distance expansions. These expansions provide two forms of solutions to the QFT model, encoding information around the infrared (IR) and ultraviolet (UV) fixed points, respectively, of the renormalization group trajectory. In parallel to the theory of differential equations, the solution to a QFT model is determined up to ``integration constants''. That is, the IR and UV expansions are {\em not} fully fixed by the QFT model: they are fixed only up to the normalizations of primary scaling fields. The IR and UV fixed points play the roles of singular points of the model, each with their own sets of QFT ``integration constants''. One of the most fundamental questions in QFT is that of the associated connection problem: setting the primary field normalizations at the UV fixed point, what are the normalizations at the IR fixed point? This contains universal information about global aspects of the renormalization group trajectory.

This connection problem is encoded into vacuum expectation values (VEVs) of primary local fields. For instance, let $\Or$ be a local spinless scaling field  normalized via the short-distance singularity of its two-point function:
\be
	\langle \Or(x)\Or(y)\rangle \sim |x-y|^{-2h_{\Or}}, \qquad
	 m |x-y| \ll 1,
\ee
where $h_{\Or}$ is the scaling dimension and $m$ is, say, the smallest mass of the massive QFT model. Then, the full correlation function $\langle \Or(x)\Or(0)\rangle$, as a function of $|x-y|$, is uniquely determined by the QFT model. In particular, by clustering, its large-distance limit is given by a product of VEVs,
\be
	\langle \Or(x)\Or(y)\rangle \sim \langle \Or\rangle^2, \qquad
	m|x-y|\gg 1,
\ee
which are then unambiguous. By dimensional analysis $\langle \Or\rangle = m^{h_{\Or}} \mathsf V_\Or$. The pure number $\mathsf V_\Or$ is a universal QFT quantity, and the set of all $\mathsf V_\Or$, for all primary local fields $\Or$, provides the full solution to the QFT connection problem. In integrable QFT, the form-factor program \cite{Kar_FF78, Smirnov_FF} provides a systematic way of obtaining the full spectral expansion, and thus very accurate correlation functions up to normalizations. However, besides exact results in important special cases \cite{Luky97VEV,VEVlu2,VEVlu3,FFLZZ}, and especially the use of an underlying fermionic symmetry algebra for VEVs of descendant fields \cite{JMS1,JMS2}, the connection problem -- the evaluation of VEVs of primary fields -- has received no systematic solution yet within integrable QFT.

Particularly interesting are {\em twist fields}:  fields $\TG_G(x)$ associated to elements $G$ of an internal symmetry group $\grp G$. Twist fields have originally been introduced in the context of the Ising model \cite{ZubI77,ST78,McCoy_76}, with the symmetry $\grp G=\mathbb{Z}_2$. They have been studied in quite some generality in CFT, where they give rise to twisted modules for vertex operator algebras \cite{dolan1990conformal, DongMason1} (also known as orbifold constructions \cite{Dixon, Verlinde89, Borisov98_orb}) by the operator-state correspondence. Correlation functions of twist fields in massive free-fermion models are related to tau functions of isomonodromic deformation problems \cite{SMJ_79,SMJ_79IV,JMU81,JM_81,McCoy_76,MTW77,Palmer1990,Palmer_94}, and the QFT connection problem for twist fields is associated to connection problems for Painlev\'e equations \cite{J_82,D03_Dirac,Lis08,Lis09,GIL13}.

A fascinating application of the concept of twist fields is to the calculation of measures of entanglement: functions of quantum states that provide a numerical indication of the quantity of entanglement in the state (see for instance \cite{Plenio05_revEE} for a review).  Two of these measures have found efficient expressions in many-body systems: the entanglement entropy (EE) \cite{Bennet_EE96}, and the logarithmic negativity (LN) \cite{Plenio_LN05,Vidal_LN02,Zyc_LN98,AEPW_LN02,E2006_phdth}. The EE was first introduced in the context of black hole entropy \cite{BKLS_86,S_93EE,CW94_EE}.   
It was first evaluated in CFT in \cite{Holzhey94}, and later  studied in the context of quantum chains \cite{AEPW_LN02,VLRK_03,LRV_04,LLRV_05,JK_04} and QFT \cite{CalabCardy_EE04,Casini05_EE,Casini05_EE2,CCD1} (see also the reviews \cite{calab09entanglement,CD09,Casini09_EE}).   
In evaluating these measures (EE or LN), a key step is the use of the replica trick \cite{CW94_EE,Holzhey94,Casini05_EE}, through which the entanglement entropy is related to partition functions on branched surfaces. In the replica trick, one is led to consider $n$ independent copies of the underlying QFT model, and there are twist fields associated to the symmetry under permutation of the copies. These twist fields were first identified as tools for the evaluation of the EE in QFT in \cite{CCD1}, where they were dubbed \emph{branch-point twist fields}. Their connection to the LN was later discovered in \cite{CalabCardy12_LN1, CalabCardy13_LN2}. Their correlation functions are related to partition functions on branched surfaces, much studied early on in CFT \cite{Dixon, bershadsky87, bershadsky88, knizhnik87,verlinde87, dijkgraaf88c} and more recently thanks to the connection with EE \cite{calab09entanglement, CCT1_2009, CCT2_2011}. In massive QFT, the form factor program for branch-point twist fields was developed and used to evaluate the EE in \cite{CCD1, CD09}, and recently to evaluate the LN in \cite{BCD1}. Of particular interest, as discovered in \cite{CCD1} and \cite{BCD1}, VEV of branch-point twist fields in massive QFT give rise to universal saturation constants for the entanglement entropy and logarithmic negativity near critical points.

Our main results are as follows. When $\grp G$ is a Lie group that is unitarily represented, and $G$ lies in the exponential map of the associated Lie algebra, $\TG_G(x)$ can be constructed from the Noether current related to the symmetry.  We provide a general formula for VEV of twist fields which can be expressed in such a way.  The formula involves the two-point function of the twist field with the Noether current, and takes into account the conformal normalization explicitly. Using a spectral expansion, it can be recast into an infinite series involving form factors of these two fields.  The series truncates to the two-particle contribution in any model of free particles, and thus gives closed expressions in such cases. In interacting integrable QFT, where form factors are known, it gives an explicit infinite-series expression for the VEV.  We expect that (generically) it is more efficient than the formula of Babujian and Karowski  \cite{BabKarow} coming from a direct form factor resummation of the logarithm of two-point functions.  In this respect, the formula we propose is in the same spirit as the infinite form-factor series representation of the conformal dimension given by Delfino, Simonetti and Cardy \cite{delfino1996asymptotic}.  

We provide explicit examples for {$\grp U(1)$} twist fields in two free-particle models: the massive Dirac fermion, and the massive complex Klein-Gordon boson. In the former, we verify that the formula agrees with known results \cite{D03_Dirac, DS_TcorrFF11}. In the latter, using a precise regularization scheme, we show that the twist field requires logarithmic renormalization, and we specify the required CFT normalization. We note that the logarithmic renormalization, in the massless case, was overlooked in the original work \cite{CW94_EE}. We show that the logarithmic renormalization is {\em different} in the massless and massive cases, and as a consequence we argue that the two-point function displays a logarithmic factor at short distances. Similar logarithmic factors were first argued for from observation of form factor divergencies in \cite{BC16_BranchP}. Under this normalization, we obtain the exact VEV of the twist field for the first time. We verify the result by using (and further explaining) the angular quantization techniques developed in \cite{Zamo_Unpub,Luky97VEV, Luky95, BazLuky98, Khoroshkin99}. In free models, branch-point twist fields can be expressed as products of $\grp{U}(1)$ twist fields. Using this, we  propose a universal saturation constant in the one-dimensional Klein-Gordon theory.

The paper is organized as follows. In Section \ref{sectoverview}, we provide an overview of the general concept of twist field in QFT, with emphasis on its realization in massive QFT. In Section \ref{sectdiffeq}, we derive our main result, a differential equation for VEV of twist fields expressed, involving its correlation function with the associated Noether current. We derive the explicit two-particle approximation, which is exact in free models. We develop the examples of the $\grp U(1)$ twist fields in the free massive Dirac  fermion, where we observe agreement with previous results, and in the free massive complex boson (a new result). In Section \ref{QuantAn}, we discuss the logarithmic renormalization of twist fields in non-compact models, and we provide an independent calculation of the VEV in the free massive complex boson that uses techniques of angular quantization. We clarify the technique by explaining in detail how to correctly introduce the conformal normalization. Finally, in Section \ref{sectE}, we apply the results to entanglement saturation via the connection between $\grp U(1)$ twist fields and branch-point twist fields.

\section{Overview of twist fields in QFT}\label{sectoverview}

In this section we provide a general overview of the concept of twist fields in QFT.   
We use Euclidean coordinate  $x=(x_1,x_2)\in \RR^2$, where $x_1$ is the space coordinate and $x_2$ the time coordinate (the formulation is entirely similar in Minkowsky space-time). 
Let $\grp{G}$ be a group of internal symmetry transformations in a 1+1-dimensional QFT model. For every group element  $G\in \grp{G}$, there exists a family $\FTG_G$ of local fields referred to as {\em twist fields} associated to $G$.    
 One characterization of this family arises naturally in the operator representation of quantization on the line. In this quantization scheme, the family $\FTG_G$ can be characterized by the following equal-time exchange relations:
\be\label{exch}
	\TG(x) \Or(y) = \left\{\begin{array}{ll} \Or(y) \TG(x) & (x_1>y_1) \\
	(G\cdot \Or)(y) \TG(x) & (x_1<y_1) \end{array}\right.,
	\qquad 
	\TG \in \FTG_G,
	\qquad x_2=y_2.
\ee
Here $\Or$ is any other local field, with $G\cdot\Or$ the result of the action of the symmetry transformation $G \in \grp G$ on $\Or$. The locality of twist fields, in the fundamental QFT sense, follows: $G$, being an internal symmetry, preserves the stress-energy tensor, wherefore twist fields in $\FTG_G$ commute at space-like distances with the stress-energy tensor.

Two immediate observations can be made from the defining relation \eqref{exch}.  For two elements $\TG_1\in \FTG_{G_1}, \TG_2\in \FTG_{G_2}$ with $G_1,G_2\in \grp G$, 
the equal-time operator product expansion (OPE), in the quantization on the line, has the form
\be\label{opeTG1TG2}
\TG_1(x) \TG_2(y) = \sum_{\TG \in \FTG_{G_1G_2}} \mathrm{C}_{\TG_1, \TG_2}^{\TG} (x-y) \TG(y)
\ee
where $ \mathrm{C}_{\TG_1, \TG_2}^{\TG}(x-y)$ are structure functions and the sum runs over a basis of $\FTG_{G_1G_2}$.  
 Moreover, the action of $G_1$ on an element $\TG_2$ is described by
\be
G_1 \cdot \FTG_{G_2^{\vphantom{-1}}}   \subseteq \FTG_{G_1^{\vphantom{-1}} G_2^{\vphantom{-1} }G_1^{-1}}.
\ee
 
As a consequence, twist fields in $\FTG_G$ form a module for ${\rm Z}_G\times \grp S$, where ${\rm Z}_G$ is the centralizer of $G$ in $\grp G$ and $\grp S$ is the space-time symmetry group.  
Since internal and space-time symmetries commute, the subspace $\FTG_G^{\grp S}\subset \FTG_G^{\vphantom{\grp S}}$ of fields that are primary under $\grp S$, is a module for ${\rm Z}_G$.     
Thus, any primary twist field in $\FTG_G^{\grp S}$ is associated to the group element $G$ and to a module element for ${\rm Z}_G$.  We will assume that within each family $\FTG_G^{\grp S}$, there is a unique primary twist field invariant under ${\rm Z}_G$, and will denote such element by the symbol $\T_G$.  That is, for each $G\in \grp{G}$, there is a unique ${\rm Z}_G$-invariant primary twist field $\T_G$.

In the following, we will be (mostly) interested in the cases when the group of internal symmetry $\grp G$ is given by (a unitary representation of) a Lie group.   
{Let $\alg g$ be the corresponding Lie algebra, and assume $G\in\grp G$ is in the image of the exponential map: it is the exponential of an element $g\in \alg g$ in the Lie algebra.  
The exchange relation \eqref{exch} suggests that the primary twist field $\T_G$ has the following formal expression:}
\be\label{formTexpJ}
\T_\alpha(x) = \T_{G^\alpha}(x_1,x_2) \propto \exp \Bigl[ 2 \pi \ii \alpha \int_{x_1}^\infty J(s, x_2) \, \dd s
\Bigr]
 , \qquad \alpha \in \RR, \qquad \ii=\sqrt{-1},
\ee
where $J(x)$ is a component of the hermitian Noether current associated to $g$ (that is, $J(y)=J_1(y)$, where $J_\nu(y) =j^\mu(y)\epsilon_{\mu\nu}$ is dual to the Noether current $j^\mu(y)$). Indeed, the exchange relation follows from the use of the Ward-Takahashi identity,
\be\label{WT}
	[J(x),\Or(y)] = (2\pi \ii)^{-1}  \delta_g \Or(y)  \,  \delta(x_1-y_1),\qquad x_2=y_2
\ee
where $ \delta_g \Or$ denote the infinitesimal action on the local field $\Or$.  The dual element will be written
\be
\bigl(\T_\alpha(x)\bigr)^\dagger = \T_{-\alpha}(x).
\ee
{Clearly, combining equal-time exchange relations for $\T_{-\alpha}(x)\Or(z)$ and $\T_\alpha(y)\Or(z)$ produces a nontrivial transformation of $\Or(z)$ only for $z_1\in[y_1,x_1]$ (if $x_1>y_1$), and thus}
\be\label{OPE2ptTTno1}
\T_{-\alpha}(x) \T_\alpha(y) \propto \exp \Bigl[ 2 \pi \ii \alpha \int_{y_1}^{x_1} J(s,y_2)  \dd s\Bigr], \qquad x_1>y_1,\  x_2=y_2.
\ee

Note that in Equations \eqref{formTexpJ} and \eqref{OPE2ptTTno1}, the twist field is defined up to a proportionality constant. This constant would be typically divergent -- the exponential on the right-hand side of \eqref{formTexpJ} usually does not converge in QFT -- and the precise definition of such twist fields, including their VEV, requires a regularization. We provide precise regularization schemes below and in Section \ref{QuantAn}.
 \\

Although the above characterizations was based on quantization on the line, twist fields can be represented in any other formulation of QFT. 
In the Feynman path-integral formulation, one simply replaces operator ordering with time ordering, as usual. The exchange relation \eqref{exch} then translates into the statement that the positions of twist fields are vertices bounding segments through which other local fields acquire discontinuities in the time direction. 
For instance, in a model with Euclidean action $S[\varphi]$, a correlation function involving a single twist field factor $\TG(x)\in\FTG_G$ and any local field or product of local fields $\Or[\varphi]$ can be represented by a path integral
\be\label{TO}
	\langle \TG(x) \Or(y) \rangle = \int_{\mathcal{C}} [\dd\varphi] e^{- S[\varphi]}
	\Or[\varphi],
\ee
where the integration is performed over all the configurations $\mathcal C$ of fields $[\varphi]$ which satisfy the following discontinuity relation {on the half-line} $\{ (s, x_2) \colon s>x_1\}$
\be\label{disc}
\mathcal C \; \colon \;  \lim_{\epsilon \to 0}\varphi(s,x_2-\epsilon) = \lim_{\epsilon\to0}(G\cdot\varphi)(s,x_2+\epsilon), \qquad s >x_1.
\ee
Different fields $\TG(x)$ in $\FTG_G$ will be associated with different asymptotic conditions on $\varphi$ near $x$. Since local fields are functionals $\Or = \Or[\varphi]$, with $\Or[G\cdot\varphi] = G\cdot \Or[\varphi]$, this translates into a discontinuity relation for the local field $\Or$ in agreement with \eqref{exch}.  
The loci of discontinuities in the path-integral formulation may be interpreted as the loci of the operators $J(s,x_2)$ in the exponential representation \eqref{formTexpJ}; however the path integral representation is more general, as 
%
it does not require $G$ to lie in the image of the exponential map of a Lie algebra.
\\

Thanks to the conservation equation $\partial_\mu j^\mu=0$, both the integration contour in expression \eqref{formTexpJ} and the loci of field discontinuities \eqref{disc} in the path-integral formulation may be deformed to any curve joining the same boundary points $x$ and $\infty$. This is true inside correlation functions, as long as the curve does not cross, under deformation, the positions of other local fields in the correlation functions. The fact that the twist field does not depend on the shape of the contour is often interpreted as an expression of its locality. In the exponential representation, the integration, on an arbitrary contour, is of the normal current component,
\be\label{Tcontour}
	\T_{\alpha}(x) \propto
	\exp\left[2\pi \ii \alpha\int_{y=x}^\infty  \sum_{\nu}
	J_\nu(y)\dd y_\nu \right].
\ee

Consider a local fields $\Or(y).$ Clearly, due to \eqref{disc}, the function $\langle \TG(x)\Or(y)\rangle$ in \eqref{TO}, as a function of $y$, is discontinuous on $\{ (s, x_2) \colon s>x_1\}$. By contour deformation as explained above, there is a unique way of continuing the function through this cut. This therefore defines a continuous function on a many-sheeted surface. With this point of view, the multi-valued function $\langle \TG(x)\Or(y)\rangle$ satisfies the following monodromy property for $y$ going around $x$:
\be
	\langle \TG(x)\,\Or(y[2\pi])\rangle =
	\langle \TG(x)\,(G\cdot\Or)(y[0])\rangle,\qquad
	y[\theta] = x + e^{\ii \theta}(y-x)
\ee
where on the left-hand side we mean the continuation from $\theta=0$ to $\theta=2\pi$ of $\langle \TG(x)\,\Or(y[\theta])\rangle$, where we identify $x$ and $y$ with their associated complex coordinates $x_1 + \ii x_2$ and $y_1 + \ii y_2$, respectively. This of course generalizes to multiple insertions of local fields.
\\

Twist fields also have a clear meaning in other quantization schemes. In radial quantization, they are related to orbifold constructions \cite{Dixon, Borisov98_orb}. We will discuss their representation in angular quantization in Section \ref{QuantAn}.

\section{VEV of twist field via differential equation}\label{sectdiffeq}

In this section, we obtain a differential equation for the vacuum expectation value (VEV) of the twist field $\T_\alpha$. Let $h_\alpha$ be the scaling dimension of the field $\T_\alpha$. We will show that
\be\label{Eq_diff_QFTno3}
\frac{1}{\pi \ii } \, \frac{\partial}{\partial \alpha} \log \langle \T_\alpha \rangle =  
\lim_{\epsilon \rightarrow 0} \Bigl( \int_{\epsilon}^{\infty}  \, \frac{ \langle  \T_{\alpha}(0)  J(s) \rangle  }{   \langle  \T_{\alpha}  \rangle    }\dd s   +  \int^{-\epsilon}_{-\infty}  \, \frac{ \langle J(s)  \T_{-\alpha}(0)  \rangle  }{   \langle  \T_{\alpha}  \rangle    } \dd s -
   \frac{1}{\pi \ii} \, \frac{\partial h_\alpha}{\partial \alpha} \log \epsilon   \Bigr)
\ee
where the right-hand side has a finite limit (all fields are at time 0). This is valid under a slightly more general short-distance normalization than that discussed in the introduction, which accounts for possible logarithmic contributions that are independent of $\alpha$: two-point functions behave, as $m|x-y|\to0$, as
\be\label{CFTnormlog2}
\langle \T_{-\alpha}(x) \T_\alpha(y)\rangle \sim  (\log(m|x-y|))^{-\ell}\,\vert x-y\vert^{-2h_\alpha}, \qquad\mbox{($\ell$ independent of $\alpha$).}
\ee
Normally one would expect $\ell=0$, but as we discuss in Section \ref{QuantAn}, logarithmic contributions occur, for instance, in models with non-compact target spaces such as the free boson theory.

{The expression given in \eqref{Eq_diff_QFTno3} is in terms of quantities that do not depend on any QFT ``integration constants'': it does not depend on normalizations of primary fields. Indeed, the right-hand side only requires the normalization of the current $J(s)$, fixed by symmetries. This expression makes clear the interpretation of the VEV as a global property of renormalization trajectories: it involves an integration from short distances ($s=\epsilon$) to large distances ($s=\infty$), along with an appropriate renormalization process ($\epsilon\to0$).}

{
Equation \eqref{Eq_diff_QFTno3} provides an expression of the VEV up to an $\alpha$-independent normalization. Indeed the VEV is obtained by integrating on $\alpha$ the right-hand side of \eqref{Eq_diff_QFTno3}, the integration constant corresponding to an overall normalization of the VEV. Under the usual CFT normalization, \eqref{CFTnormlog2} with $\ell=0$, the integration constant is fixed by requiring
\be\label{normell0}
	\lim_{\alpha\to0}\langle \T_\alpha\rangle = 1 \qquad(\ell=0).
\ee
This requirement arises because at $\alpha=0$, the field $\T_\alpha$ is the identity field. On the other hand, if $\ell\neq0$ in \eqref{CFTnormlog2}, then it is clear that the limit $\alpha\to0$ cannot be uniform: the short-distance behavior, in the limit $\alpha=0$, still has logarithmic contributions, while the identity field does not have such contributions. In this case, we do not know of a universal way of connecting the integration constant arising after integration over $\alpha$, to the $\alpha$-independent normalization \eqref{CFTnormlog2}. We will fix the integration constant according to a definite renormalization scheme in the complex free boson theory in Section \ref{QuantAn}.
}

\subsection{Derivation of \eqref{Eq_diff_QFTno3}}

 Recall that a VEV is, essentially, a property of two-point functions. We therefore use the representation \eqref{OPE2ptTTno1} of two-point functions, and exploit the fact that $\alpha$ is a continuous parameter.  Let us define
\be\label{Falpha}
F_\alpha = F_\alpha(\vert x_1-y_1\vert)= \langle \T_{-\alpha}(x) \T_{\alpha}(y) \rangle = A_\alpha \langle   \exp \Bigl( 2 \pi \ii \alpha \int_{y_1}^{x_1} J(s,y_2)  \dd s\Bigr) \rangle, \qquad(x_2=y_2).
\ee
{Here we have explicitly introduced a formal UV-divergent constant $A_\alpha$.} The derivative of $F_\alpha$ with respect to $\alpha$ gives the relation 
\be \label{eq_diff_1_NR}
\frac{1}{2\pi \ii } \, \frac{\partial}{\partial \alpha} \log F_\alpha = \int_{y_1}^{x_1}  \, \frac{ \langle  J(s,y_2) \T_{-\alpha}(x) \T_{\alpha}(y) \rangle  }{   \langle  \T_{-\alpha}(x) \T_{\alpha}(y) \rangle   } \, \dd s   
+ \frac1{2\pi \ii} \frac{\partial}{\partial\alpha} \log A_\alpha.
\ee
Observe that this last equation expresses the value of a two-point function in terms of an integration of a three-point function containing the symmetry current $J(s,y_2)$. { In general, the integral in the right-hand side of \eqref{eq_diff_1_NR} is ill-defined because of divergencies at colliding positions, as $s$ approaches either $x_1$ or $y_1$. It thus needs to be UV regularized. A natural regularization is the replacement of the integration limits by
\be\label{regul}
	x_1\mapsto x_1-\epsilon, \quad y_1\mapsto y_1+\epsilon
\ee
for some small $\epsilon>0$. Choosing appropriately $A_\alpha=A_\alpha(\epsilon)$ in order to cancel the divergency, this regularization can then be sent away, $\epsilon\to0$, and one obtains a renormalized, finite two-point function. The appropriate choice can be obtained by using the CFT normalization, as follows.  }

The conformal behavior of primary twist fields is in general of the form
\be\label{CFTnormlog1}
F_\alpha =  \vert x_1-y_1\vert^{-2h_\alpha} \qquad \mbox{(CFT)}
\ee
where $h_\alpha$ is the scaling dimension of $\T_\alpha$. This is to be understood as the two-point function of twist fields evaluated in the CFT corresponding to the UV fixed point of the model. Taking the derivative, we have
\be\label{danormacft1}
 \frac{\partial}{\partial \alpha} \log F_\alpha = -2 \, \frac{\partial h_\alpha}{\partial \alpha} \, \log \vert x_1-y_1 \vert \qquad \mbox{(CFT)}.
\ee

{In general, CFT is expected to reproduce the short-distance behavior of correlation functions, $m\vert x_1-y_1\vert \ll1$, and thus \eqref{CFTnormlog1} should describe the short-distance behavior of \eqref{Falpha}. However, as discussed in Section \ref{QuantAn}, in models with non-compact target spaces the relation between the CFT behavior and the short-distance limit may be more subtle, because of potential logarithmic contributions to the renormalization procedure. It will be shown explicitly that, in the case of the free complex Klein-Gordon field, the same field necessitates different logarithmic renormalization procedures in the massive and massless models. In this case, although \eqref{CFTnormlog1} is the correct CFT two-point function, the difference in renormalization procedures indicates that the short-distance behavior of \eqref{Falpha} in the massive model has additional logarithmic factors as per \eqref{CFTnormlog2}. Such factors might be interpreted as coming from logarithmic contributions at higher orders in conformal perturbation theory (although we will not discuss this particular aspect). Nevertheless, as will be confirmed in the Klein-Gordon model, the extra logarithmic factors in the renormalization procedures are not expected to depend upon $\alpha$. As a consequence, $\p \log A_\alpha(\ep)\,/\,\p\alpha$ in \eqref{eq_diff_1_NR} is the same in the CFT as in the massive model.}

By a general CFT argument  (related to the Knizhnik-Zamolodchikov equation), the three-point function has the following form
\be\label{trepointJconsv}
\frac{ \langle J(s,y_2)  \T_{-\alpha}(x) \T_{\alpha}(y) \rangle  }{   \langle  \T_{-\alpha}(x) \T_{\alpha}(y) \rangle   } = \frac{d_1}{s-y_1} + \frac{d_2}{x_1-s}, \qquad y_1 < s < x_1 \qquad\mbox{(CFT).}
\ee
The constants $d_1=d_1(\alpha)$ and $d_2=d_2(\alpha)$ depend on $\alpha$.  Going back to equation \eqref{eq_diff_1_NR},   integration over $s$ of \eqref{trepointJconsv} yields logarithmic divergences whenever $s$ approach $x_1$ or $y_1$.  Using the regularization \eqref{regul}, the constant $A_\alpha=A_\alpha(\epsilon)$ is therefore fixed in terms of $d_1$ and $d_2$,
\be
	\frac1{2\pi \ii}
	\frac{\partial}{\partial\alpha} \log A_\alpha = (d_1+d_2) \log\epsilon.
\ee
Performing the regularized integral in \eqref{eq_diff_1_NR} in the CFT using \eqref{trepointJconsv}, and comparing with \eqref{danormacft1}, this further fixes uniquely
\be\label{d1plusd2}
d_1+d_2 = -\frac{1}{\pi \ii} \, \frac{\partial h_\alpha}{\partial \alpha}.
\ee
Therefore, the renormalized differential equation \eqref{eq_diff_1_NR}, valid for all distances $x_1-y_1$, is
\be\label{eq_diff_RE2}
\frac{1}{2\pi \ii} \, \frac{\partial}{\partial \alpha} \log F_\alpha =  \lim_{\epsilon \rightarrow 0} \Bigl( \int_{y_1+\epsilon}^{x_1-\epsilon}  \, \frac{ \langle  J(s,y_2) \T_{-\alpha}(x) \T_{\alpha}(y) \rangle  }{   \langle  \T_{-\alpha}(x) \T_{\alpha}(y) \rangle    }  \, \dd s   -\frac{1}{\pi \ii} \, \frac{\partial h_\alpha}{\partial \alpha} \log \epsilon \Bigr).
\ee
The right-hand side of this equation is indeed well-defined.

In order to finally obtain the differential equation for the expectation value $\langle \T_\alpha \rangle$, we consider equation \eqref{eq_diff_RE2} at large distances, for instance $y_1=0$ and $x_1=\infty$.  By the clustering property,  $F_\alpha\rightarrow (\langle \T_\alpha \rangle)^2$, and by a rescaling we obtain \eqref{Eq_diff_QFTno3}.
 
\subsection{Two-particle approximation}

The two-point function $\langle \T_\alpha(0)J(s)\rangle$ and its complex conjugate $\langle J(s) \T_{-\alpha}(0)\rangle$ can be expanded with the aid of form factors.  \ Consider the following representation of the identity (i.e. completeness relation)
\be
{\bf 1} = \sum_{k=0}^\infty \; \sum_{\mathsf j} \;\int_{-\infty}^{\infty} |\ta_1\ldots \ta_k; \mathsf j \rangle \, \langle \ta_1 \ldots \ta_k; \mathsf j | \, \frac{\dd \ta_1 \ldots \dd \ta_k}{(2\pi)^k k!}
\ee
where the $\ta_i$ variables are the rapidity and $\mathsf j=(\mathsf j_1, \ldots, \mathsf j_k)$ represents the $k$-vectors of eigenvalues (or quantum numbers) associated to the state $\vert \ta_1\ldots \ta_k;\mathsf j\rangle$ under the symmetry described by $J(s)$.%
\footnote{In the following, we will use twist fields that are associated to the $\grp U(1)$ group, so that the eigenvalue $\mathsf j_i$ of the charge will take values in $\{\pm 1\}$.} {We will denote by $|{\mathrm{vac}}\rangle$ the vacuum vector (with $k=0$).}

{Let us assume for simplicity that there is no particle type with zero quantum number (that is, invariant under $G^\alpha$ for all $\alpha$).} Inserting the completeness relation in the correlation function, {no one-particle contribution occurs and we find}
  \be\label{exp_ff_no1_a}
\frac{ \langle  \T_{\alpha}(0)  J(s) \rangle  }{   \langle  \T_{\alpha}  \rangle    } =
\sum_{\mathsf j} \int_{-\infty}^\infty  \frac{  \langle \mathrm{vac} | \T_\alpha(0)|\theta_1 \theta_2  ; \mathsf j  \rangle  \, \langle \ta_1 \ta_2; \mathsf j \vert J(s) \vert \mathrm{vac}\rangle}{    \langle  \T_{\alpha}  \rangle   } \, \frac{\dd \ta_1 \dd \ta_2}{2 (2\pi)^2} + \ldots
\ee
Here we have used the fact that $\langle J(s)\rangle=0$. The matrix elements involved are called form factors, and they are exactly known in integrable QFT.  For interacting integrable QFT, the first two-particle term in the  expression \eqref{exp_ff_no1_a} are expected to provide a good approximation.  For free theory, the field  $J(s)$ has nonzero form factors only at $2$-particle order, and thus the terms explicitly written in \eqref{exp_ff_no1_a} give an exact expression.

Let us introduce the variables
\be\label{cov_pm12}
\tau_1= \tfrac12(\ta_1-\ta_2), \qquad \tau_2 = \tfrac12(\ta_1+\ta_2), \qquad \dd\ta_1 \dd\ta_2 = 2 \dd \tau_1 \dd \tau_2,
\ee
and denote the matrix elements by
\begin{align}\label{FFT}
&\langle \mathrm{vac} | \T_\alpha(0)|\theta_1 \theta_2  ; \mathsf j \rangle = \langle \T_\alpha \rangle  \; \mathcal F^{(\T_\alpha)}(\tau_1; \mathsf j), \\
&\langle \mathrm{vac} | J(0)|\theta_1 \theta_2  ; \mathsf j \rangle =  \mathcal F^{(J)}(\tau_1,\tau_2; \mathsf j) .
\end{align}
Note that by translation invariance, we have $\langle \mathrm{vac} | J(s)|\theta_1 \theta_2  ; \mathsf j \rangle =
\exp\bigl( 2\ii  m  s \cosh\tau_1 \sinh\tau_2 \bigr)\;\mathcal F^{(J)}(\tau_1,\tau_2; \mathsf j)$.
In \eqref{FFT} we have used the fact that $\T_\alpha(0)$ is a scalar under Lorentz transformations (i.e. rotations, in the Euclidean formulation) in order to write its form factor as a function of the difference of the rapidities.

Let us evaluate the first integral in \eqref{Eq_diff_QFTno3} in the two-particle approximation:
 \be\label{EQ_Diff_ok_fin3} 
 \int_{\epsilon}^{\infty}  \, \frac{ \langle  \T_{\alpha}(0)  J(s) \rangle  }{   \langle  \T_{\alpha}  \rangle    }\dd s = 
 \int_\epsilon^\infty \dd s \;   
 \int_{-\infty}^{+\infty}   \frac{\dd \tau_1 \dd \tau_2}{(2\pi)^2} \;
  e^{ -2\ii  m  s \cosh\tau_1 \sinh\tau_2 }
  \sum_{\mathsf j}  \,
\mathcal F^{(\T_\alpha)}(\tau_1; \mathsf j) (\mathcal F^{(J)}(\tau_1, \tau_2; \mathsf j)
)^* .
\ee
We make the following change of variable $\tau_2 \mapsto \tau_2-\ii\pi/2$.  This results in changing the integration over $\tau_2$, as
\be
\int_{-\infty}^\infty\dd \tau_2 \mapsto \int_{-\infty + \ii\pi/2}^{\infty+\ii\pi/2}\dd \tau_2
\ee
and we can chose to close the integration contour as
\be
\int_{-\infty +\ii\pi/2}^{\infty+\ii\pi/2}\dd \tau_2 + \int_{\infty + \ii\pi/2}^{\infty}\dd \tau_2 +\int_{\infty}^{-\infty}\dd \tau_2 + \int_{-\infty}^{-\infty +\ii\pi/2}\dd \tau_2 = \oint_{\mathrm C} \dd \tau_2=0
\ee
since there is no poles inside $\mathrm C$. The second and last integration give a zero contribution. Using $\sinh(\tau_2 - \ii\pi/2) = -\ii\cosh\tau_2$, we may integrate the result over $s$,  and \eqref{EQ_Diff_ok_fin3} becomes
\be
 \int_{\epsilon}^{\infty}  \, \frac{ \langle  \T_{\alpha}(0)  J(s) \rangle  }{   \langle  \T_{\alpha}  \rangle    }\dd s = 
 \int_{-\infty}^{+\infty}   \frac{\dd \tau_1 \dd \tau_2}{(2\pi)^2} \;
  \frac{e^{-2m\epsilon \cosh\tau_1\cosh\tau_2}}{2m\cosh\tau_1\cosh\tau_2}
  \sum_{\mathsf j}  \,
\mathcal F^{(\T_\alpha)}(\tau_1; \mathsf j) \big(\mathcal F^{(J)}\big)^*(\tau_1, \tau_2-\ii\pi/2; \mathsf j)
\ee

Then the solution of equation \eqref{Eq_diff_QFTno3} is, up to two-particle terms, written as
\be \label{EQ_Diff_ok_fin1} 
  \langle \T_\alpha \rangle =  \mathrm  N \exp \Bigl\{ \pi \ii \int 
 \,  W (\alpha) \, \dd \alpha \Bigr\}
 \ee
where
 \be\label{EQ_Diff_ok_fin2}  \begin{split}
 W(\alpha) & = 
 \lim_{\epsilon \rightarrow 0} \Bigl \{ 
 \int_{-\infty}^{+\infty}   \frac{\dd \tau_1 \dd \tau_2}{(2\pi)^2} \;
  \frac{e^{-2m\epsilon \cosh\tau_1\cosh\tau_2}}{2m\cosh\tau_1\cosh\tau_2}  \sum_{\mathsf j}  \,
       \Bigl[
\mathcal F^{(\T_\alpha)}(\tau_1; \mathsf j) \big(\mathcal F^{(J)}\big)^*(\tau_1, \tau_2-\ii\pi/2; \mathsf j)
\\
& \qquad \qquad  \qquad
+ \mathcal F^{(J)}(\tau_1,\tau_2-\ii\pi/2; \mathsf j) \big(\mathcal F^{(\T_{-\alpha})}\big)^*(\tau_1; \mathsf j) \Bigr]
 -
  \frac{1}{\pi \ii} \, \frac{\partial h_\alpha}{\partial \alpha}  \log (\epsilon) \Bigr\}.
   \end{split}
\ee

The normalization constant $\mathrm N$ {comes from} the integration over $\alpha$. As mentioned and as we will see in the example of the Dirac fermion, in models with compact target spaces the VEV is expected to be fully fixed by \eqref{normell0}.  On the other hand, for non-compact target spaces such as in the uncompactified free boson, then we expect the solution \eqref{EQ_Diff_ok_fin2} to have a divergency as $\alpha\to0$ and the VEV cannot be fixed in this way. This is  discussed in Section \ref{QuantAn}.

We now provide explicit calculations in two examples where equation \eqref{EQ_Diff_ok_fin2} can be evaluated exactly (the massive Dirac fermion and the massive complex boson). In both examples, the symmetry group is $\grp{G} = \grp{U}(1)$, and we choose $\alpha\in[0,2\pi)$ to represent the phase in the fundamental representation of $\grp U(1)$.

\subsection{The Dirac fermion}
The simplest case where we can illustrate the general method presented in this section for the  computation of VEV is the free massive Dirac fermion.  
VEVs of $\grp U(1)$ twist fields in this model were first studied in the scheme of the angular quantization (which will be the subject of Section \ref{QuantAn}) in  \cite{Luky97VEV, BazLuky98}, see also \cite{D03_Dirac, DS_TcorrFF11}. Note that the $\grp{U}(1)$ current field in the Dirac model is, after bosonizing, given by a sine-Gordon boson, hence the twist field is an exponential field in the sine-Gordon theory. This specializes to a free massless boson vertex operator in the CFT limit.

Let us first recall the known result. Let $\langle \T_\alpha \rangle = m^{h_\alpha} \mathsf V_\alpha$ be the one-point function of the $\grp U(1)$ twist field for the Dirac fermion theory where $m$ represent the mass and $h_\alpha=\alpha^2$ is the scaling dimension of $\T_\alpha$.  The value of $\mathsf V_\alpha$ is given by
\be\label{vev_know_GG_int1}
\mathsf V_\alpha = 
\frac{2^{-\alpha^2}}{G(1+\alpha) G(1-\alpha)} = 
2^{-\alpha^2}\exp  \Bigl(\,  \int_0^\infty  \Bigl[ \frac{\sinh^2 (\alpha t )}{\sinh^2 t} - \alpha^2  e^{-2 t} \Bigr] \frac{\dd t}{t} \,  \Bigr),
\ee
where the function $G(\cdot)$ is the Barne $G$-function. \\

We now show that \eqref{Eq_diff_QFTno3} reproduces this result. The theory of a Dirac fermion (complex fermion) is constructed from two real fermions $\psi_1(x)=\psi_1(x_1,x_2)$ and $\psi_2(x)=\psi_2(x_1,x_2)$ with  $x$ the Euclidean coordinate.     
We write $\psi=(\psi_1+\ii \psi_2)/\sqrt 2$, and   also  $\psi^\dagger=(\psi_1-\ii \psi_2)/\sqrt 2$, and denote by $m$ the parameter usually associated to the mass.  
The field satisfies the non-trivial commutation relation
\be \label{psipsi12} \psi(x_1,y) \psi^\dagger(x_2,y) + \psi^\dagger(x_2,y) \psi(x_1,y) =  4\pi \delta(x_1-x_2).
 \ee
 There is an internal $\grp U(1)$ symmetry for the free complex fermion theory since the action is invariant under re-parametrization $\psi\mapsto e^{\ii \alpha} \psi$.  The associated current $J(x)$ is
 \be
 J(x) = (4\pi)^{-1}\normord{\psi(x)\psi^\dagger(x)}
 \ee
 where $\normord{\Or}$ denotes normal order.  Setting
 \be
 \T_\alpha(x) \propto  \exp \Bigl[ \frac{\ii \alpha}{2} \int_x^\infty \psi(s) \psi^\dagger(s) \dd s \Bigr], 
 \ee
  one can verify that the relation \eqref{exch} with $G\cdot \psi= e^{2 \pi \ii \alpha} \psi$ is satisfied, using the anti-commutation relation \eqref{psipsi12}. One can also verify easily that the CFT three-point function \eqref{trepointJconsv} has the form given by the equivalent computation of a correlation function of vertex operators, which are known exactly.

We now present the computation of function $W(\alpha)$, defined in \eqref{EQ_Diff_ok_fin2}.  The matrix elements (i.e.~the two-particle form factors) can be calculated by standard methods \cite{Kar_FF78,SMJ_79,SMJ_79IV,Smirnov_FF,BernLeC94}, and are
\be\begin{split}
\mathcal F^{(\T_\alpha)}(\tau_1;+-)&={\ii} \sin(\pi \alpha) \frac{e^{2 \alpha \tau_1}}{\cosh \tau_1}  \\
 \mathcal F^{(\T_\alpha)}(\tau_1;-+) &= {-\ii}\,  \sin(\pi \alpha) \frac{e^{-2 \alpha \tau_1}}{\cosh \tau_1}  
\end{split}
\ee
where $+-, -+$ denotes the eigenvalues $\mathsf j\in (\mathbb Z_2)^2$ of $J(x)$.  Matrix elements associated to the others eigenvalues $++,--$ vanish.  It is straightforward to  obtain 
\be
\mathcal F^{(J)}(\tau_1, \tau_2;+-) = -\, \mathcal F^{(J)}(\tau_1, \tau_2;-+)=
-\ii m e^{  \tau_2}.
\ee
Plugging these into \eqref{EQ_Diff_ok_fin2}, we obtain
 \be\label{EQ_Diff_ok_fin6}  \begin{split}
 W(\alpha) & = 
 \lim_{\epsilon \rightarrow 0} \Bigl \{ \ii\sin\pi\alpha
 \int_{-\infty}^{+\infty}   \frac{\dd \tau_1 \dd \tau_2}{2\pi^2} \;e^{-2m\epsilon \cosh\tau_1\cosh\tau_2}
  \frac{\cosh(2\alpha\tau_1)}{\cosh^2\tau_1}  
 +
  \frac{2 \ii \alpha }{\pi } \, \log (\epsilon) \Bigr\}\\
  &=  \lim_{\epsilon \rightarrow 0} \Bigl \{ \ii\sin\pi\alpha
 \int_{-\infty}^{+\infty}   \frac{\dd \tau_1 }{\pi^2} \;K_0(2m\epsilon \cosh\tau_1)
  \frac{\cosh(2\alpha\tau_1)}{\cosh^2\tau_1}  
 +
 \frac{2 \ii \alpha }{\pi } \, \log (\epsilon) \Bigr\}
   \end{split}
\ee
where $K_0(\cdot)$ is the modified Bessel function. For small values of $z$, the Bessel function is given by $K_0(z) \sim -\gamma_{\mathrm E}-\log(z/2)$ where $\gamma_{\mathrm E}$ denotes the Euler constant. We therefore obtain
\be
\begin{split}
W(\alpha) &= \lim_{\epsilon \rightarrow 0} \Bigl(  \frac{-2\ii }{\pi^2} \sin(\pi \alpha) \int_0^\infty \frac{\cosh(2\alpha \tau)}{\cosh^2 \tau} \log(\cosh \tau) \dd \tau\\
&\qquad   - \frac{2\ii }{\pi^2} (\gamma_{\mathrm E}+ \log(m \epsilon))     \sin(\pi \alpha)  \int_0^\infty \frac{\cosh(2\alpha \tau)}{\cosh^2 \tau} \dd \tau  + \frac{2\ii \alpha}{\pi} \log(\epsilon) \Bigr).
\end{split}
\ee
The divergency in $\log(\epsilon)$ in the previous expression cancels exactly thanks to
\be
\int_0^\infty \frac{\cosh(2\alpha \tau)}{\cosh^2 \tau} \dd \tau = \frac{\pi \alpha}{\sin(\pi \alpha)} .
\ee
Since the $\tau$ integral converges uniformly over any compact subset of $(-1,1)$, the integration over $\alpha$ can be performed. Fixing the integration constant such that $\langle \T_{0}\rangle=1$, we obtain the result
\be
\langle \T_\alpha \rangle =m^{\alpha^2} \exp\Big( \gamma_{\mathrm E} \alpha^2+ \frac{2}{\pi} \int_0^\infty \frac{\log(\cosh \tau)}{(\pi^2 + 4\tau^2) \cosh^2 \tau } \bigl[ \pi - \pi \cos( \pi \alpha) \cosh(2\alpha \tau) + 2 \tau \sin(\pi \alpha ) \sinh(2\alpha \tau) \bigr] \dd \tau \Bigr).
\ee
Writing $\langle \T_\alpha\rangle = m^{\alpha^2} \mathsf V_\alpha$ as above,  one can compare this result numerically with the expression \eqref{vev_know_GG_int1} for all $\alpha\in (0,1)$, and we have observed perfect agreement.

\subsection{The complex Klein-Gordon boson}
\label{tiboso}
We now present the calculation for the VEV of the twist field in a theory of a free complex boson.  This twist field was first studied in \cite{SMJ_79IV}, but its VEV was never evaluated.  We will break the calculation into intermediate steps. First, we set the notation in the quantization on the line. Second, we study the conformal field theory of the $\grp U(1)$ twist field and its current. This is in order to explicitly establish, in the free boson case, the relations \eqref{trepointJconsv} and \eqref{d1plusd2}; recall that this is the part of the general theory at the basis of the correct normalization for the VEV. Here, one important difference with the Dirac fermion is that the use of vertex operator for the calculation of correlators is not possible since the current $J(s)$ is typically a bilinear expression in the free bosonic modes (it cannot be expressed in linear bosonic form). Finally, we will use these ingredients in the general theory that we have established in order to obtain an expression for the twist field VEV. We will see that there is a divergency when $\alpha\rightarrow 0$. This is connected with the logarithmic structure discussed in Section \ref{QuantAn}.

\subsubsection{Setting the notation}  The free massive complex  (Klein-Gordon) boson is made of two independent real free bosonic fields  $\phi_1=\phi_1(x_1,x_2)$ and $\phi_2=\phi_2(x_1,x_2)$, from which we build a single complex free bosonic field $\phi=\phi_{1}+\ii \phi_{2}$.  
Its conjugate is denoted by $ \bphi=\phi_{1}-\ii \phi_{2}$.  The Klein-Gordon boson theory is described by the following Hamiltonian
\be\label{HmfreeBo1}
H=\int_{-\infty}^\infty  \normord{ \Pi^\dagger \Pi + ( \partial_1 \bphi) (\partial_1 \phi) + m^2 \bphi \phi } \dd x, \qquad \Pi={\ii \partial_2 \phi}
\ee
where $m$ is the mass, and $\partial_i$ denotes partial derivative w.r.t.~$x_i$.  Recall that we are using Euclidean coordinates with imaginary time $x_2$, for instance $(\partial_2)^\dagger = - \partial_2$.    
  The normal ordering is, as usual, with respect to the vacuum $|0\rangle$ of the Fock module of the mode algebra. The mode expansion is
\be\label{fcb_phi}
\phi(x_1,x_2) = \frac{1}{\sqrt{4\pi}} \int_{-\infty}^{\infty} \bigl( e^{- \ii p_\ta x_1+ \ii E_\ta x_2} A_+^\dagger(\ta)
+ e^{\ii p_\ta x_1- \ii E_\ta x_2} A_-^{\phantom \dagger}(\ta) \bigr) \dd \ta, \qquad p_\ta=m \sinh \ta, \quad E_\ta=m \cosh \ta
\ee
where $A_\pm^{\phantom \dagger}(\ta), A_\pm^{\dagger}(\ta)$ are operators that satisfy $[A_\pm^{\phantom \dagger}(\ta_1), A_\pm^{\dagger}(\ta_2)]=\delta(\ta_1-\ta_2)$,  other commutators vanishing. The vacuum is defined by $A_\pm(\ta) |0\rangle =0$ for all $\ta$, and the associated normal ordering is
\be
\normord{A_\pm^{\phantom \dagger}(\ta_1) A_\pm^\dagger(\ta_2)} = A_\pm^\dagger(\ta_2)A_\pm^{\phantom \dagger}(\ta_1) .
\ee
The (canonical) commutation relations for the complex boson are
\be\label{fcb_crel_Piphi}
[\bphi(x_1,x_2), \Pi(y_1,y_2) ] = [\phi(x_1,x_2), \Pi^\dagger(y_1,y_2) ]= \ii \delta(x_1-y_1), \qquad x_2=y_2.
\ee
The complex boson theory has an internal $\grp U(1)$ symmetry, and a state with a set of rapidities $\ta_k$ is also characterized by the associated $\grp U(1)$ charges $\mathsf j_k=\pm1$.  Such a state is written as $|\ta_1,\ldots,\ta_n;\mathsf j_1,\ldots,\mathsf j_n \rangle=A^\dagger_{\mathsf j_1}(\ta_1)\cdots A^\dagger_{\mathsf j_n}(\ta_n)|0\rangle$.  

It is a simple matter to show that the $\grp U(1)$ symmetry current $J(x)$ is
\be\label{U1currJfcb1}
J = \ii \normord{\Pi^\dagger \phi - \bphi \Pi}.
\ee
The 2-particle form factors 
 can be obtained explicitly using the mode expansion \eqref{fcb_phi}  and the commutation relations, and are
\be\label{FFJ_fcb1}\begin{split}
\mathcal F^{(J)}(\tau_1, \tau_2; +-) &= - \,  \mathcal F^{(J)}(\tau_1, \tau_2; -+)= \frac{1}{4\pi}(E_{\ta_1}-E_{\ta_2})  \\
&= \frac{m}{2\pi} \sinh \tau_1 \sinh\tau_2,
\end{split}
\ee
the other combinations being zero (recall the variables $\tau_1$ and $\tau_2$ in \eqref{cov_pm12}).

The current $J(x)$ \eqref{U1currJfcb1} enters the exponential expression of the twist field as per \eqref{formTexpJ}. With the aid of the commutation relations \eqref{fcb_crel_Piphi}, one immediately sees that this representation for the twist field satisfies the relation \eqref{exch} with $G\cdot \phi = e^{2\pi \ii \alpha} \phi$.  Its scaling dimension is known the be \cite{Dixon}
 \be\label{halphaboson}
h_\alpha= \alpha(1-\alpha).
\ee

The two-particle form factors for the twist fields can be evaluated by solving the form factor equations, and are given by \cite{SMJ_79IV}
\be \begin{split}
\mathcal F^{(\T_\alpha)}(\tau_1;+-) &= {2\pi}\sin(\pi \alpha) \frac{   e^{(2\alpha-1)\tau_1} }{\cosh\tau_1 }   \\
\mathcal F^{(\T_\alpha)}(\tau_1;-+) &= {2\pi}\sin(\pi \alpha) \frac{   e^{-(2\alpha-1)\tau_1} }{\cosh\tau_1 }
 .
  \end{split}
\ee

\subsubsection{The CFT normalization}  A key step in the derivation of the differential equation for the VEV of $\T_\alpha$ is the evaluation of the CFT three-point function
  \be\label{traviata}
 \frac{\langle \T_{-\alpha}(x) J(s) \T_\alpha(0) \rangle}{ \langle \T_{-\alpha}(x) \T_\alpha(0) \rangle}.
 \ee  
We will obtain in this section that this expression is in agreement with the general decomposition given in \eqref{trepointJconsv}, with the explicit value \eqref{d1plusd2} (with \eqref{halphaboson}).

Consider the Hamiltonian \eqref{HmfreeBo1} with $m=0$, and introduce the complex variables $z,\bar z$ given by $z=x_1+\ii x_2, \bar z=x_1-\ii x_2$ so that $\phi(x_1,x_2)$ becomes a function of complex variables, $\phi(z,\bar z)$. Below we will drop the explicit dependence on the anti-holomorphic variable $\bar z$. The partial holomorphic and anti-holomorphic derivatives are denoted by $\partial=\partial_z= \tfrac12 (\partial_1-\ii\partial_2)$ and $\bar\partial = \partial_{\bar z} =\tfrac12  (\partial_1 +\ii \partial_2)$.    The corresponding CFT has central charge $c=2$, and the fields $\partial \phi$ and $\partial \phi^\dagger$ are two primary fields with singular OPE
\be
\partial \phi^\dagger(z) \partial \phi(w) \sim \frac{-1}{4\pi} \, \frac{1}{(z-w)^2},
\ee
in agreement with the commutation relations \eqref{fcb_crel_Piphi}. Similar relations hold for anti-holomorphic derivatives of $\phi$.   Recall that $\partial \phi$ and $\partial \phi^\dagger$ are holomorphic fields, and $\bar\partial \phi(z)$ and $\bar\partial \phi^\dagger(z)$ are anti-holomorphic fields.

In term of complex coordinates, the $\grp{U}(1)$ current field is written 
\be\label{fcb_Jzbzfacto}
J (z) =
-\ii \bigl(\normord{\partial \bphi(z) \phi(z)}  - \normord{\bphi(z) \partial \phi(z)}\bigr)  \, + \, \ii \bigl(\normord{\bar \partial \bphi(z) \phi(z)}  - \normord{\bphi(z) \bar \partial \phi(z)}\bigr).
\ee
Observe that the current decouples into two similar parts, which we treat independently.  Define $J_1(z)$ ($J_2(z)$) to be the part of $J(z)$ that only contains holomorphic (anti-holomorphic) derivatives.  Thus, the expression \eqref{traviata} breaks into 
\be
 \frac{\langle \T_{-\alpha}(x) J_1(z) \T_\alpha(0) \rangle}{ \langle \T_{-\alpha}(x) \T_\alpha(0) \rangle} +  \frac{\langle \T_{-\alpha}(x) J_2(z) \T_\alpha(0) \rangle}{ \langle \T_{-\alpha}(x) \T_\alpha(0) \rangle}.
\ee
Each term can be evaluated using the radial quantization scheme, where we can interpret these correlators as expectation values of $J_1(z)$ (or $J_2(z)$) in a twisted module of the free-field algebra.

Let us set temporally $\alpha$ to some rational value $\alpha=k/n$ for $n\in\mathbb Z_{>1}$ and $k \in \{ 0,1,\ldots, n-1\}$.  The field $\T_{k/n}(0)$ transforms the radial-quantization vacuum $\vert 0\rangle$ (with radial-quantization center at the origin) into the highest weight of a twisted module, to be denoted by $\T_{k/n}(0)\vert 0 \rangle = \vert \sigma_k\rangle$.  Such modules for CFT are well known and have been studied in the context of orbifold   \cite{Dixon, DongMason1}.     
In the twisted module generated by $\vert \sigma_k\rangle$, the field $\partial \phi$ can be expanded as
\be\label{vertexfieldphiH1}
\partial\phi(z) = \sum_{m\in \mathbb Z} a_{m-k/n} \, z^{-(m-k/n)-1}
\ee
and the field $\partial \phi^\dagger$ as
\be
\partial\phi^\dagger(z) = \sum_{m\in \mathbb Z} a^\dagger_{m+k/n} \, z^{-(m+k/n)-1}.
\ee
The Laurent modes satisfy the relation $[a_{m+k/n}^{ \dagger}, a_{m'-k/n}^{\phantom \dagger}] = \mathrm g\, (m+k/n) \delta_{m,-m'} {\bf 1}$ where $\mathrm g$ is the constant given by the normalization of the fields, i.e. $\mathrm g=(-4\pi)^{-1}$. The twisted vacuum is defined by the relations
\be
	a_{m-k/n} \vert \sigma_k \rangle = 0 \quad (m>0),\qquad
	a^\dagger_{m+k/n} \vert \sigma_k\rangle\quad (m\geq 0).
\ee

The dual vectors correspond to fields at infinity and we will denote $\langle 0 \vert \T_{-k/n}(\infty)=\langle \sigma_{-k}|$.   
By direct computation, we find
\be \label{cordpdpb1} \begin{split}
\frac{\langle \sigma_{-k} | \partial \phi^\dagger(z) \partial \phi(w) |\sigma_k \rangle }{\langle \sigma_{-k} | \sigma_k \rangle} &
= \frac{1}{\langle \sigma_{-k}| \sigma_k \rangle} \sum_{m,m' \in \mathbb Z}
 \langle \sigma_{-k}| a^\dagger_{m+k/n}  a_{m'-k/n}^{\phantom \dagger} |\sigma_k \rangle z^{-(m+k/n)-1} w^{-(m'-k/n)-1} \\
& = \mathrm{g} \sum_{m\geq 0}(m+k/n) z^{-m-k/n-1}w^{m+k/n-1} \\
&=  \mathrm{g}\, z^{-k/n} \partial_w  \Bigl(  \frac{w^{k/n}}{z-w}   \Bigr) \\
& =  \mathrm{g}\, z^{-k/n}w^{k/n-1} \Bigl( \frac{ kz/n + w(1-k/n)}{(z-w)^2} \Bigr).
\end{split}
\ee
As $w\rightarrow z$, we see that this last result contains a singularity of the form $(z-w)^{-2}$. The normal-ordered field $\normord{\partial \phi^\dagger \partial \phi}$, where normal ordering is with respect to the vacuum $|0\rangle$, is obtained by removing this singularity:
\be
\frac{\langle \sigma_{-k} |  \normord{\partial \phi^\dagger(z)\partial \phi(z)} |\sigma_k \rangle }{\langle \sigma_{-k} | \sigma_k \rangle}
=
\lim_{w\rightarrow z} \Bigl( \frac{\langle \sigma_{-k} | \partial \phi^\dagger(z) \partial \phi(w) |\sigma_k \rangle }{\langle \sigma_{-k} | \sigma_k \rangle} - \frac{\mathrm{g}}{(z-w)^2}\Bigr) = \frac{\mathrm{g}}{2}\frac{ (k/n) (1-k/n)}{z^2}.
\ee
Note that this method is sometimes referred to as the point-splitting technique.  Observe that since the stress-energy tensor is  $T(z)= \mathrm{g}^{-1}  \normord{\partial\phi^\dagger(z) \partial \phi(z)}$, we can deduce the scaling dimension $h_\alpha$ of the twist field from this last equation, by replacing each occurence of $k/n$ with $\alpha$, in agreement with \eqref{halphaboson}.

Under the assumption that the twist fields are primary, the fields that appear in the correlator in \eqref{cordpdpb1} are all primary fields. Therefore doing the conformal transformation
\be
z\; \mapsto \; u(z)=\frac{z}{x-z}
\ee
will result into the relation
\be\label{expvvsexpsigma}
\frac{\langle  \T_{-k/n}(x) \partial \phi^\dagger(z) \partial \phi(w) \T_{k/n}(0) \rangle}{\langle  \T_{-k/n}(x) \T_{k/n}(0) \rangle}   =  \Bigl( \frac{\dd u(z)}{\dd z}\Bigr) \, \Bigl(\frac{\dd u(w)}{\dd w} \Bigr) \,  \frac{\langle \sigma_{-k} | \partial \phi^\dagger(u(z)) \partial  \phi(u(w)) |\sigma_k \rangle }{\langle \sigma_{-k} | \sigma_k \rangle}
\ee
since $\partial \phi$ is of conformal dimension $1$.   
Integrating \eqref{expvvsexpsigma} with respect to $w$, this gives
\be
\frac{\langle  \T_{-\alpha}(x) \partial \phi^\dagger(z) \phi(w) \T_\alpha(0) \rangle}{\langle  \T_{-\alpha}(x) \T_\alpha(0) \rangle}  =
 \mathrm{g}\, w  \Bigl(\frac{z}{x-z}\Bigr)^{-\alpha} 
\, \Bigl( \frac{w}{x-w}\Bigr)^{\alpha-1} \, \frac{1}{(z-w)(x-z)}.
\ee
Using the point-splitting technique, i.e. removing the singularity of the form $(z-w)^{-1}$ when $w\rightarrow z$, we have
\be
\frac{\langle  \T_{-\alpha}(x) \normord{\partial \phi^\dagger(z) \phi(z)} \T_\alpha(0) \rangle}{\langle  \T_{-\alpha}(x) \T_\alpha(0) \rangle}  =
 \lim_{w\rightarrow z} \Bigl(  \frac{\langle  \T_{-\alpha}(x) \partial \phi^\dagger(z) \phi(w) \T_\alpha(0) \rangle}{\langle  \T_{-\alpha}(x) \T_\alpha(0) \rangle} -  \frac{\mathrm{g}}{z-w} \Bigr) =
  \frac{\mathrm{g}\, (z-\alpha x)}{z(x-z)}.
\ee
This gives us the first contribution that appears in $J_1(z)$.  
  By a similar technique, we obtain the second contribution of $J_1(z)$ from
\be
\frac{\langle  \T_{-\alpha}(x) \normord{ \phi^\dagger(z)  \partial \phi(z)} \T_\alpha(0) \rangle}{\langle  \T_{-\alpha}(x) \T_\alpha(0) \rangle}  
= \lim_{w\rightarrow z} \Bigl(  \frac{\langle  \T_{-\alpha}(x)  \phi^\dagger(w) \partial \phi(z) \T_\alpha(0) \rangle}{\langle  \T_{-\alpha}(x) \T_\alpha(0) \rangle} -  \frac{\mathrm{g}}{z-w} \Bigr) =  \frac{\mathrm{g}\, (\alpha x -x+z)}{z(x-z)}.
\ee
Combining the two previous results, we find
\be\label{nouv_eq_J1_expV}
\frac{\langle \T_{-\alpha}(x) J_1(z) \T_\alpha(0) \rangle}{ \langle \T_{-\alpha}(x) \T_\alpha(0) \rangle}
 = -\ii \Bigl( \frac{-1}{4\pi } \Bigr) 
 \Bigr[  \frac{z-\alpha x}{z(x-z)}   -  \frac{\alpha x-x+z}{z(x-z)}      \Bigr] = \frac{1}{4\pi \ii} \, \frac{-x(1-2\alpha)}{z(x-z)}.
\ee
The equivalent expression for $J_2(z)$ gives exactly the same result, so the total three-point function \eqref{traviata} is just given by a factor $2$ times \eqref{nouv_eq_J1_expV}. This is indeed of the form \eqref{trepointJconsv}, and it is easy to check that \eqref{d1plusd2} indeed holds with \eqref{halphaboson}. In particular, the integration is then
\be\label{regucftfbno1}
 \int_\epsilon^{x-\epsilon} \frac{\langle \T_{-\alpha}(x) J(z) \T_\alpha(0) \rangle}{ \langle \T_{-\alpha}(x) \T_\alpha(0) \rangle}  \dd z
 =  - \frac{1}{\pi \ii} \, {(1-2\alpha)} \log x + \frac{1}{\pi \ii}(1-2\alpha) \log \epsilon
\ee
where we see that the logarithm divergency as $\epsilon\rightarrow 0$ is exactly cancelled by 
\be
-\frac{1}{\pi \ii} \partial_\alpha (h_\alpha) \log \epsilon
\ee
in agreement with the general expression \eqref{eq_diff_RE2} and \eqref{d1plusd2}.

\subsubsection{VEV} 
We end by giving the explicit expression for the VEV of the twist field for the free massive complex boson.    
As with the Dirac fermion case, using the corresponding form factors, we have for $W(\alpha)$
  \be\label{eq_sol_pr_1}\begin{split}
W(\alpha) &= 
\lim_{\epsilon\rightarrow 0} \Bigl( \frac{2\ii}{\pi^2} \sin(\pi \alpha) \int_0^\infty 
\frac{\sinh(\tau) \sinh((2\alpha-1)\tau)}{\cosh^2(\tau)} \, \bigl[ \gamma_{\mathrm E}+ \log(m\epsilon) + \log(\cosh \tau) \bigr] \, \dd \tau  \\
& \qquad \qquad \qquad  
- \frac{1}{\pi \ii} (1-2\alpha) \log(\epsilon) \Bigr).
\end{split}
\ee
With the formula
\be
\int_0^\infty \frac{\sinh(\tau) \sinh((2\alpha-1)\tau)}{\cosh^2(\tau)} \dd \tau = \frac{\pi}{2} \frac{(2\alpha-1)}{\sin(\pi \alpha)},
\ee
we find that the divergency in \eqref{eq_sol_pr_1} cancels out and we have
\be
W(\alpha) = \frac{1}{\pi \ii} (1-2\alpha) ( \gamma_{\mathrm E} + \log(m)) - \frac{2}{\pi \ii} \sin (\pi \alpha)  \, \int_0^\infty \frac{\sinh(\tau) \sinh((2\alpha-1)\tau)}{\cosh^2(\tau)} \, \log(\cosh\tau) \, \dd \tau.
\ee
Integrating over $\alpha$ finally gives
\be\begin{aligned}
 &\langle \T_\alpha \rangle 
 = \mathrm N\, (me^{\gamma_{\mathrm E}})^{\alpha (1-\alpha)}\, \times \\ &
\quad\times \exp \Bigl( - \frac2\pi  \int_0^\infty 
 \frac{(\sinh \tau) \log( \cosh \tau ) }{   (\pi^2 + 4 \tau^2)   \cosh^2 \tau } \, 
 \bigl[ \pi \cos(\pi \alpha) \sinh((1-2\alpha) \tau) + 2 \tau \sin( \pi \alpha) \cosh((1-2\alpha) \tau)\bigr] \dd \tau \Bigr)
 \end{aligned}\label{VEVbos}
\ee
where $\mathrm N$ is an integration constant.

Here, when $\alpha\rightarrow 0$, the value of the VEV goes to zero proportionally to $\sqrt{\alpha}$, and thus the constant $\mathrm N$ cannot be fixed by requiring the VEV to be unity in this limit (this requirement is incorrect).   This is associated with a logarithm behavior in the two-point function of twist field, which we make more precise in the next section.

We remark that comparing with the result of the next section, the correct normalization is given by
\be\label{Ncst}
\log \mathrm N = - \frac13 ( \log(2) + \gamma_{\rm E}) + \frac12(1 - \log(2\pi)) - \int_0^\infty \frac{\dd \tau}{\tau} e^{-\tau} \Bigl( \frac{\sinh (\tau) -\tau}{ (\cosh(\tau)-1) (e^{-\tau}-1)} + \frac13 \Bigr).
\ee

\section{VEV of twist fields and their logarithmic structure from  angular quantization}
\label{QuantAn}

The goal of this section is two-fold. First, we provide a definition of regularized twist fields, and discuss how logarithmic factors are involved in the renormalization process when the target space of the QFT is non-compact. This renormalization scheme is particularly well adapted to the techniques of angular quantization. We explain how angular quantization may be used in order to evaluate, under this renormalization scheme, two-point functions of twist fields in CFT, and, following \cite{Luky97VEV, BazLuky98}, VEVs of twist fields in massive free QFT. This allows us to make an explicit connection between the CFT normalization and the VEV, an essential ingredient in the evaluation of the VEV that was overlooked in \cite{Luky97VEV}.

Second, we apply these concepts to the complex uncompactified free boson. We obtain the following alternative expression for the VEV:
\be\label{VEVboson}
	\langle \T_\alpha\rangle =
	\lt(\frac{m}2\rt)^{h_\alpha}\,\sqrt{\frac{\sin( \pi \alpha)}\pi}\,\exp\lt[
	\int_0^\infty e^{-t}
	\Big(\frac{2\sinh (t\alpha/2)\,\sinh (t(1-\alpha)/2)}{(1-e^{-t})\sinh (t/2)}-\alpha(1-\alpha)\Big) \frac{\dd t}{t} 
	\rt].
\ee
We have verified numerically that this agrees with the result \eqref{VEVbos} under the choice \eqref{Ncst}. This thus confirms the method developed in the previous section. We also explicitly show how logarithmic factors arise in the renormalization of the $\grp U(1)$ twist fields, both in the CFT calculation and in the massive QFT calculation. In particular, with $\T_\alpha^\ep(x)$ a natural UV-regularized twist field (see below), we find in CFT that, as the distance-like regularization $\ep$ is sent to zero,
\be\label{Zexpov}
\langle \T_{-\alpha}^\epsilon(0)\T_\alpha^\epsilon(x)\rangle_{\rm CFT}  \propto |\log(\ep/x)|\,\lt|\frac{\epsilon}x\rt|^{2h_\alpha},
\ee
and in the massive case, we find
\be\label{Zexpovmass}
\langle \T_\alpha^\epsilon\rangle  \propto |\log(m\ep)|\,\lt|m\epsilon\rt|^{h_\alpha}.
\ee
We argue that the short-distance limit of the massive two-point function of renormalized fields also has  a logarithmic factor as follows:
\be\label{ttov}
	\langle \T_{-\alpha}(0)\T_\alpha(x)\rangle \sim
	|\log(m|x|)|^{-1}\,|x|^{-2h_\alpha}
	\qquad(m|x|\to0).
\ee
Note how the powers of the logarithms are different in the last two equations. The fact that the power of the logarithm is negative in \eqref{ttov} is in agreement with recent numerical form factor observations \cite{BC16_BranchP} (although the exact power is in disagreement with that conjectured there).

\subsection{Regularizations of twist fields and logarithmic renormalization in non-compact models} \label{ssectreg}

As mentioned in Section \ref{sectoverview}, the exponential definition of twist fields \eqref{formTexpJ} and \eqref{OPE2ptTTno1} requires an appropriate regularization and renormalization procedure. In Section \ref{sectdiffeq}, a regularization was obtained, after differentiation with respect to $\alpha$, by point splitting, \eqref{Falpha} and \eqref{regul}. There are many other ways of regularizing the exponential, and all should give rise to the same universal results.

In order to extract more accurately the structure of UV divergencies of the twist field, and to verify the technique developed in Section \ref{sectdiffeq} for the evaluation of VEVs, in this section we study a natural geometric regularization. 
We define a regularized twist field $\T_\alpha^\ep$, with distance-like regularization parameter $\ep>0$, as follows. Recall that the exponential form of twist fields leads, in the path-integral formulation, to a definition in terms of field discontinuities, \eqref{TO} and \eqref{disc}. The regularized twist field is defined by this path-integral formula, but where the space $\RR^2$, on which the integration field $\varphi$ lives, is missing a disk of radius $\ep>0$ around the position of the twist field. On the boundary of this disk, an appropriate boundary condition is imposed. We choose the boundary condition to be uniform, and such that the twist field has smallest dimension possible (so that it is primary); generically, this corresponds to imposing some low-divergency uniform asymptotic of the path-integral field $\varphi$ near the insertion point of the twist field. The exact boundary condition (or asymptotics) depends on the model, and will be made more precise below in the context of the massive free boson. Many boundary conditions may be chosen, and one expects these to give rise to all twist fields in the family $\FTG_{G^\alpha}$.

The renormalization process is the taking of the limit $\ep\to0$, after appropriate renormalization factors have been introduced. The boundary conditions imposed near the positions of the twist fields usually reduce the space of field configurations and thus the value of the path integral. Therefore these renormalization factors are chosen to diverge, in such a way that the resulting correlation functions are finite. By locality of the QFT, each field insertion gives its own divergent factor, and thus is associated to an independent renormalization constant. The precise divergency depends on the model, and a generically power-like divergency gives rise to a nontrivial, positive scaling dimension (equal to this power).\\

In order to illustrate the definition, let us consider the one-point functions (VEVs) and two-point functions. We have
\be\label{ta1}
	\langle\T_\alpha^\ep\rangle
	= \frac{\int_{\mathcal{C}_\alpha(\ep,\infty)} \Big[
	\dd\varphi|_{\RR^2\setminus D_\ep(0)}\Big]{ e^{-S[\varphi]}}}{
	\int_{\mathcal{C}_0(\ep,\infty)} \Big[
	\dd\varphi|_{\RR^2\setminus D_\ep(0)}\Big]{ e^{-S[\varphi]}}}
\ee
and
\be\label{ta2}
	\langle \T_{-\alpha}^\ep(0) \T_\alpha^\ep(x)\rangle
	= \frac{\int_{\mathcal{C}_\alpha(\ep,x_1-\ep)} \Big[
	\dd\varphi|_{\RR^2\setminus (D_\ep(0) \cup D_\ep(x))}\Big]{ e^{-S[\varphi]}}}{
	\int_{\mathcal{C}_0(\ep,x_1-\ep)} \Big[
	\dd\varphi|_{\RR^2\setminus (D_\ep(0) \cup D_\ep(x))}\Big]{ e^{-S[\varphi]}}}
\ee
where (recall that $g$ is the Lie algebra element associated to $\T_\alpha$)
\be\label{disc3}
\mathcal C_\alpha(a,b) \; \colon \;  \varphi(s,x_2-0^+) = (e^{2\pi \ii \alpha g}\cdot\varphi)(s,x_2+0^+), \qquad s \in[a,b]
\ee
and where $D_\ep(x)=\{z\in\CC:|z-x|<\ep\}$ is a disk of radius $\ep$ centred at $x$. Note that the integration field $\varphi$ of the path integral is defined on the Euclidean space with, respectively, one and two hole(s), $\RR^2\setminus D_\ep(0)$ and $\RR^2\setminus (D_\ep(0)\cup D_\ep(x))$, and that in the numerator the integration field has a discontinuity on a cut starting at $\ep$, and in the denominator it is continuous everywhere.

For primary fields, the usual procedure for obtaining universal correlation functions is to take the limit on $\ep$ renormalized by a product of powers of $\ep$, one for each twist field insertion, corresponding to the dimensions $h_\alpha$ of the twist fields inserted:
\be\label{nolog}
	\langle\T_{\alpha_1}(x_1)\cdots \T_{\alpha_n}(x_n)\rangle
	= \Bigl(\prod_j \mathrm c_{\alpha_j}\Bigr) \,\lim_{\ep\to0} \ep^{-\sum_j
	h_{\alpha_j}}
	\langle\T_{\alpha_1}^\ep(x_1)\cdots \T_{\alpha_n}^\ep(x_n)\rangle.
\ee
The normalization constants $\mathrm c_\alpha$ are, as explained in the introduction, the ``integration constants'' of QFT, and may be chosen in such a way that the CFT two-point function has the form\footnote{Imposing the VEV to be real, we may choose $\mathrm c_\alpha = \mathrm c_{-\alpha}\in \RR$ and $h_\alpha=h_{-\alpha}\in\RR$.} \eqref{CFTnormlog1}. With this choice, the VEV is a well-defined universal quantity pertaining to the QFT model.

The renormalization \eqref{nolog} is indeed correct in most situations. However, there are situations where the power-like divergency does not renormalize correctly the field (for instance the result is infinite). We will argue that in models with non-compact target space (where local degrees of freedom take values in a non-compact space\footnote{More precisely, the topology of the target space is determined by the local interaction between the degrees of freedom, this local interaction being the source of the target-space metric.}), this prescription must generically be modified.

The qualitative argument is as follows. Non-compactness of the target space implies that fluctuations of the path-integral fields may be locally very large. Indeed, the field configurations in the path integral are almost surely continuous, but not differentiable; non-compactness of the target space therefore allows for large field fluctuations (in an operator language, these are zero-mode effects). This has the consequence that at near the disk-like holes of radii $\ep$, there may be additional contributions to the path integral, that gives it a higher value than expected. The power-law regularization is thus too strong; generically, one must divide by powers of logarithms, $|\log \ep|^\ell$, for some $\ell>0$. Calculations in the free boson models will show that such powers do not depend upon $\alpha$, and we will assume so in the following. 
However, contrary to the power-law renormalizations, there is {\em not necessarily} an independent logarithmic contribution for each field insertion. Indeed, such logarithmic contributions  do not localize at field insertions if correlations are too strong. If the distances between twist field insertions are much smaller than the correlation length (that is, $mx\ll 1$, while of course $x\gg\ep$) then the fluctuations at each twist field insertion are so correlated that one might expect there to be a single logarithmic contribution. 
On the other hand, if the distances are much greater than the correlation length, every field insertion might give rise to its own logarithmic contribution. In the latter case, correlation functions of regularized fields cluster, and so the latter is a statement about VEVs.

Therefore we expect in general that universal correlation functions at large distances (that is, products of universal VEVs) be obtained as
\be\begin{aligned}\label{logmass}
	& \lim_{mx_j\to\infty}
	\langle\T_{\alpha_1}^\ep(x_1)\cdots \T_{\alpha_n}^\ep(x_n)\rangle = \prod_j \langle \T_{\alpha_j}^\ep\rangle\qquad\mbox{
	($m\ep$ fixed)} \\
	\mbox{and}\qquad &
	\langle \T_\alpha\rangle = \mathrm c_\alpha
	\lim_{\ep\to0} \frac{\ep^{-h_\alpha}}{|\log\ep|^{\ell_1}}
	\langle \T_\alpha^\ep\rangle,
	\end{aligned}
\ee
while in the CFT limit (with $n>1$),
\be\begin{aligned}\label{logCFT}
	& \lim_{mx_j\to0}
	\langle\T_{\alpha_1}^\ep(x_1)\cdots \T_{\alpha_n}^\ep(x_n)\rangle =
	\langle\T_{\alpha_1}^\ep(x_1)\cdots \T_{\alpha_n}^\ep(x_n)\rangle_{\rm CFT}
	\qquad\mbox{
	($x_j/\ep$ fixed)} \\
	\mbox{and}\qquad &
	\langle\T_{\alpha_1}(x_1)\cdots \T_{\alpha_n}(x_n)\rangle_{\rm CFT}
	= \Bigl(\prod_j \mathrm c_{\alpha_j}\Bigr) \,\lim_{\ep\to0} \frac{\ep^{-\sum_j
	h_{\alpha_j}}}{|\log\ep|^{\ell_n}}
	\langle\T_{\alpha_1}^\ep(x_1)\cdots \T_{\alpha_n}^\ep(x_n)\rangle_{\rm CFT}
	\end{aligned}
\ee
for some $\ell_1$, $\ell_n\;(n>1)$ independent of $\alpha$. If, as argued, there is a single logarithmic contribution at short distances because of strong correlations, while there are as many independent contributions as there are field insertions at large distances, then we would expect
$	\ell_n=\ell_1$.\\

Note that the above definitions still guarantee the correct scaling behavior of the twist fields (without logarithms), as upon scaling $\ep\mapsto \lambda\ep$, the additional constant arising, $\log\ep\mapsto \log\lambda+\log\ep$, is negligible in the limit $\ep\to0$. In particular, under appropriate choices of $\mathrm c_\alpha$ the two-point function has the usual form
\be
	\langle \T_{-\alpha}(0) \T_\alpha(x)\rangle_{\rm CFT} = |x|^{-2h_\alpha}
\ee
in agreement with \eqref{CFTnormlog1}. This also indicates that the universal correlation functions are indeed independent of the particular choice of UV regularization scheme.
We make five additional comments:
\begin{itemize}
\item We have not discussed correlation functions for $mx_j$ nonzero and finite. We do not have independent calculations to verify any conjecture in this case, however, since factorization of correlation function should hold both for the (universal) renormalized fields and the (non-universal) regularized one, it is reasonable to assume that each field receives an independent logarithmic correction:
\be\label{logmassplus}
	\langle\T_{\alpha_1}(x_1)\cdots \T_{\alpha_n}(x_n)\rangle
	= \Bigl(\prod_j \mathrm c_{\alpha_j}\Bigr) \,\lim_{\ep\to0} \frac{\ep^{-\sum_j
	h_{\alpha_j}}}{|\log\ep|^{n\ell_1}}
	\langle\T_{\alpha_1}^\ep(x_1)\cdots \T_{\alpha_n}^\ep(x_n)\rangle
	\qquad (m>0).
\ee
\item It is clear from the above discussion that if $\ell_n \neq n\ell_1$, as is expected in non-compact models, the short-distance limit {\em does not commute} with the renormalization $\ep\to0$ limit. This is an effect of non-compactness, giving rise to new divergencies in $\ep$.
\item If $\ell_n \neq n\ell_1$, because of the change of logarithmic renormalization, the short-distance behavior of massive correlation functions $\langle\T_{\alpha_1}(x_1)\cdots \T_{\alpha_n}(x_n)\rangle$ are not given by the CFT correlation function as defined above. Extra logarithmic factors appear as in \eqref{CFTnormlog2}. They may be understood as logarithmic contribution at higher orders in the conformal perturbation theory. The exact power may be obtained by arguing as follows. Consider the two-point function $n=2$ in \eqref{logmassplus}. There are $2\ell_1$ powers of $|\log\ep|$ in the denominator. Yet, as the two points become nearer to each other, according to \eqref{logCFT} the correct normalization has $\ell_2$ powers. If $2\ell_1>\ell_2$, then there are too many powers in the denominator, so the short-distance two-point function becomes much smaller than the CFT two-point function. Assuming that this happens as a power of $\log(m |x|)$ that agrees with the missing power of $|\log\ep|$, we have
\be\label{aofje}
	\langle \T_{-\alpha}(0)\T_\alpha(x)\rangle \sim
	(\log(m|x|))^{\ell_2-2\ell_1}\,|x|^{-2h_\alpha}
	\qquad(m|x|\to0).
\ee
Therefore we identify in \eqref{CFTnormlog2} the power $\ell$ to be
\be\label{ell12gen}
	\ell = 2\ell_1-\ell_2.
\ee
\item As mentioned in Section \ref{sectdiffeq}, the limit $\alpha\to0$ is clearly singular if $\ell_1,\ell_n\neq 0$. Indeed, at $\alpha=0$ the twist field is the identity field, which is not affected by logarithmic renormalization, while the regularization factors still have logarithms in the denominator in the limit $\alpha\to0$. This explains why, in non-compact cases, the VEV tends to zero or infinity as $\alpha\to0$. Note that it must tend to 0, for instance, in the case $\ell_1>0$ (intuitively because we have taken away too many logarithmic factors in defining the VEV):
\be\label{Talpha0}
	\lim_{\alpha\to0} \langle\T_\alpha\rangle = 0 \qquad (\ell_1>0).
\ee
\item One may also define logarithmic twist fields by subtraction: the above extracts the leading logarithmically divergent part, corresponding to primary scaling fields, and subtracting it gives rise to universal correlation functions of logarithmic partners to the twist fields. We will not discuss such subtractions in the present paper.
\end{itemize}

\subsection{Angular-quantization formulae for VEV of twist fields}

We presented in Section \ref{sectoverview} a definition of twist fields based on quantization on the line. We also showed how this leads to its definition based on the path integral formalism. From the path integral formalism, one can then consider other quantization schemes, and obtain twist fields in these other schemes. For instance, in Subsection \ref{tiboso}, we used the implementation of twist fields as twisted modules in CFT: this is their implementation, via the field-state correspondence, in radial quantization. In the present section, we construct twist fields in the angular quantization scheme. This scheme is particularly powerful in free-particle (quadratic) models, and leads to an explicit evaluation of VEVs of twist fields in such models. This will allow us to compare with the expression in the complex Klein Gordon model obtained from the proposed form factor formula in Section \ref{sectdiffeq}, and thus confirm its validity. Angular quantization was first used in \cite{Luky97VEV, BazLuky98} in order to evaluate VEVs of twist fields, where the VEV of the $\grp U(1)$ Dirac twist field was first evaluated (and we recall that the expression obtained in Section \ref{sectdiffeq} agrees with this). Here we provide appropriate clarifications of the technique, which allow us to use it more generally.\\

In the angular quantization scheme, which is a formulation specific to the Euclidean context, imaginary time is chosen to be the angle around a center (say the origin), and space is the logarithm of the radial variable. Therefore, equal-time slices are rays emanating from the center, the angular quantization Hilbert space is formed by field configurations on the half-line (the rays), and the angular quantization Hamiltonian $H_{\rm an}$ is the operator, on that Hilbert space, that generates rotations around the center. The Hamiltonian $H_{\rm an}$ is independent of angular-quantization time whenever the theory is rotation invariant. Denoting Euclidean positions with complex variables $x=x_1+\ii x_2$, we will use the angular quantization variables
\be\label{tranAngQuan1}
	\eta+\ii\xi = \log(x)
\ee
where $\eta$ is the space, and $\xi$ is the imaginary time. Since time is cyclic, averages (with respect to the vacuum of quantization on the line, or equivalently in their path-integral expressions on Euclidean space) are represented by traces in angular quantization space,
\be\label{expValOrastraceno2}
	\langle \Or \rangle = \frac{\tr\left(e^{-2\pi H_{\rm an}}\Or\right)}{\tr\left(e^{-2\pi H_{\rm an}}\right)}.
\ee

Observe that thanks to current conservation, the field $\int_0^{\infty} J(y_1,0)\,\dd y_1$ is invariant under rotation with respect to the origin (see \eqref{Tcontour}). Since the angular quantization Hamiltonian generates rotations, this means that the representation $\mathcal{R}_{\rm an}\left[\int_0^{\infty} J(y_1,0)\,\dd y_1\right]$ of this field on the angular quantization space is an operator that commutes with $H_{\rm an}$ (here and below, $\mathcal{R}_{\rm an}\left[\Or(x)\right]$ represents the map from fields $\Or(x)$ to their representation on the angular quantization space). By definition, this is the integration over the full angular quantization space of a conserved Noether current, hence it is the (Hermitian) conserved charges $Q_{\rm an}$ associated to the corresponding internal symmetry:
\be
	Q_{\rm an}= \mathcal{R}_{\rm an}\left[\int_0^{\infty} J(y^1,0)\,\dd y^1\right],\qquad
	[Q_{\rm an},\mathcal{R}_{\rm an}\left[\Or(x)\right]] = (2\pi \ii)^{-1} \mathcal{R}_{\rm an}\left[\delta_g\Or(x)\right]
\ee
and
\be\label{QHan}
	[Q_{\rm an},H_{\rm an}] = 0.
\ee
Using \eqref{formTexpJ}, we then have, formally,
\be\label{VEVtrace0}
	\langle \T_{\alpha} \rangle \propto \frac{\tr\left(e^{-2\pi (H_{\rm an}-\ii\alpha Q_{\rm an})}\right)}{\tr\left(e^{-2\pi H_{\rm an}}\right)}.
\ee

The ratio of traces on the right-hand side of \eqref{VEVtrace0} is in general expected to be UV divergent, see for instance \cite[App.B]{Luky97VEV}. The regularization discussed in Subsection \ref{ssectreg} is well adapted to angular quantization, as it preserves rotation invariance in the case of one-point functions. In this case, the regularization restricts the $\eta$-space to the half-infinite line $\eta\in[\log\ep,\infty]$ for some small $\ep>0$, and imposes an appropriate boundary condition at $\log\ep$. The ratio of partition functions in \eqref{ta1} is therefore represented as follows in angular quantization:
\be \label{VEVtrace_eps1}
	\langle \T_\alpha^\ep\rangle =
	\frac{\tr_\ep\left(e^{-2\pi (H_{\rm an}-\ii\alpha Q_{\rm an})}\right)}{\tr_\ep\left(e^{-2\pi H_{\rm an}}\right)}
\ee
where $\tr_\ep$ is the trace on the regularized angular Hilbert space.   One can then obtain $\langle \T_\alpha\rangle$ from expression \eqref{VEVtrace_eps1} by using the general renormalization \eqref{logmass}, where recall that $h_\alpha$ is the scaling dimension of $\T_\alpha$. The traces in \eqref{VEVtrace_eps1} may be evaluated by simultaneous diagonalization of $H_{\rm an}$ and $Q_{\rm an}$.

Expression \eqref{VEVtrace_eps1}, while expected to be finite, does not however fix the VEV: the constant $\mathrm c_\alpha$ in the renormalizaion \eqref{logmass} is not yet specified. In \cite{Luky97VEV} it was conjectured that, in the free massive Dirac model, imposing appropriate conformal boundary conditions would guarantee that expression \eqref{VEVtrace_eps1} (in the compact case) gives the correct VEV (that of the twist field under CFT normalization \eqref{CFTnormlog1}) if one chooses $\mathrm c_\alpha=1$. Below we provide instead a first-principle reasoning, applicable more generally, and find that in general $c_\alpha\neq1$.

Consider the regularized CFT two-point function $\langle \T_{-\alpha}^\ep(0) \T_\alpha^\ep(x)\rangle_{\rm CFT}$ for some $x>0$. This is the ratio of partition functions \eqref{ta2}. This ratio is invariant under M\"obius transformations, as the partition functions are affected by the same factor in the numerator and denominator. Consider the transformation $z\mapsto z/(x-z)$. This maps $0$ to itself and $x$ to $\infty$. In fact,  up to corrections that become negligible as $\ep\to0$, it maps $D_\ep(0) \to D_{\ep/x}(0)$ and $D_\ep(x)\to D_{\ep/x}(\infty)=\{z\in \CC\cup\infty:|z|^{-1}<\ep/x\}$. Under this transformation, therefore, the region $\RR^2\setminus (D_\ep(0) \cup D_\ep(x))$ is mapped onto $\RR^2\setminus (D_{\ep/x}(0) \cup D_{\ep/x}(\infty))$ as $\ep\to0$, which is isometric under rotations centred at the origin. The path integral on this region can therefore be represented in the angular quantization scheme, where the $\eta$-space is the finite interval $\eta\in[\log(\ep/x),\log (x/\ep)]$. As a consequence, we have
\be\label{tpCFTtrace}
	\langle \T_{-\alpha}^\ep(0)\T_{\alpha}^\ep(x) \rangle_{\rm CFT}
	=
	\frac{\tr_{\ep/x,x/\ep}\left(e^{-2\pi (H_{\rm an}^{\rm CFT}-\ii\alpha Q_{\rm an}^{\rm CFT})}\right)}{\tr_{\ep/x,x/\ep}\left(e^{-2\pi H_{\rm an}^{\rm CFT}}\right)}
	\ee
where $\tr_{\ep,L}$ is the angular quantization trace on the finite interval $\eta\in[\log \ep,\log L]$. The renormalized fields are then evaluated using \eqref{logCFT}. Rescaling $\ep\mapsto x\ep$ and using the conformal normalization $\langle \T_{-\alpha}(0)\T_{\alpha}(x) \rangle_{\rm CFT} = x^{-2h_\alpha}$, we find
\be\label{Balpha}
	\mathrm c_\alpha = \lim_{\ep\to0} \left(\frac{
	\ep^{-2h_\alpha}}{|\log\ep|^{\ell_2}}
	\frac{\tr_{\ep,1/\ep}\left(e^{-2\pi (H_{\rm an}^{\rm CFT}-\ii\alpha Q_{\rm an}^{\rm CFT})}\right)}{\tr_{\ep,1/\ep}\left(e^{-2\pi H_{\rm an}^{\rm CFT}}\right)}\right)^{-1/2}.
\ee

Thus, expression \eqref{VEVtrace_eps1} using the renormalization \eqref{logmass}  with \eqref{Balpha} gives an exact general formula for VEVs of twist fields in the angular quantization scheme, which does not rely on any ad-hoc assumption.

\subsection{Logarithmic structure and exact VEV in the complex Klein-Gordon model}\label{ssectQuanAnKG}

The evaluation of the traces in \eqref{VEVtrace_eps1} may be particularly difficult in interacting models. However, in free-particle models, it is possible by using Fock spaces. We perform these calculations in the complex Klein-Gordon theory. We use the Dirichlet boundary condition at the boundary of the hole in the regularized definition of the twist fields. Therefore, we impose $\phi(x=\ep e^{\ii \theta})=0$ for the field at the origin, and we impose additionally $\phi(x=\ep^{-1}e^{\ii\theta})=0$ for the field at $\infty$ in the evaluation of $\mathrm c_\alpha$ (see \eqref{Balpha}). This guarantees that the field is well defined both in the massive QFT and in the CFT.

We will show the following. The renormalizations for the one-point and two-point functions in \eqref{logmass} and  \eqref{logCFT}  give finite results if and only if we set
\be\label{ell12bos}
	\ell_1=\ell_2=1.
\ee
According to \eqref{ell12gen}, this sets $\ell=1$ in the short-distance logarithmic behavior \eqref{CFTnormlog2}, and thus gives \eqref{ttov}. Further, we obtain the renormalization constant
\be\label{calphabos}
	\mathrm c_\alpha = \frac12 \sqrt{\frac{\pi}{\sin\pi\alpha}}
\ee
and we find the universal VEV \eqref{logmass} to be given by \eqref{VEVboson}.
Note that we have
\be
	\langle \T_\alpha\rangle \sim \sqrt{\alpha}, \; \qquad
	\alpha\to0,
\ee
verifying that \eqref{Talpha0} holds.

\subsubsection{Angular quantization of the model}

Using the transformation \eqref{tranAngQuan1} in the action of the complex Klein Gordon model, and performing the usual steps in order to extract the Hamiltonian, one can evaluate the angular quantization Hamiltonian explicitly in terms of fields:
\be\label{Han}
\Han=\int_{-\infty}^\infty \normord{  \Piand \Pian  + (\partial_\eta \phiand) ( \partial_\eta \phian) + m^2 e^{2\eta} \phiand \phian} \dd \eta, \qquad \Pian= \ii\partial_{\xi} \phian,
\ee
where
\be
	\phian(\eta,\xi)=  \mathcal{R}_{\rm an}[\phi(x)], \qquad \Pian(\eta,\xi)= e^\eta  \mathcal{R}_{\rm an}[\Pi(x)]
\ee 
and the equal-time canonical commutation relations hold:
\be\label{canan}
	[\phiand(\eta), \Pian(\eta') ] = [\phian(\eta), \Piand(\eta') ]= \ii \delta(\eta-\eta').
\ee
In addition, we have the following angular conserved $\grp U(1)$ charge  
\be\label{Qan}
\Qan = \ii \int_{-\infty}^\infty  \normord{   \Piand \phian - \phiand  \Pian       }  \dd \eta.
\ee 

Note that in the expression for the Hamiltonian $\Han$, the mass term has an extra $\eta$-dependent factor $e^{\eta}$ as compared to the expression for the Hamiltonian in quantization on the line \eqref{HmfreeBo1}.  This is sometimes referred to as a mass barrier, since as $\eta$ becomes large, the potential becomes large. This precludes waves from emanating to or from positive infinity, causing them rather to reflect off the mass barrier. Because of this reflection, it effectively reduces the number of degrees of freedom by half, as it should since the half-line supports half as many field configurations. It also guarantees that the field's asymptotic condition as $\eta\to\infty$, which in the angular quantization scheme corresponds to the effect of a field positioned at infinity, is always zero. This means that any field placed at infinity has no effect, as expected from large-distance exponential clustering in massive QFT. In the massless case, the mass barrier is not present, and an appropriate condition must be imposed at infinity in order to construct the angular Hilbert space. This condition corresponds to a field placed at infinity, whose effect still is important; one then obtains expressions for two-point functions, \eqref{tpCFTtrace}.

Under the regularization, the integrals in \eqref{Han} and \eqref{Qan} run over the appropriate subregions of the line.

\subsubsection{The regularized CFT two-point function}\label{s432}
Consider the case $m=0$, and the expression \eqref{Balpha}. The traces $\tr_{\ep/x,x/\ep}$ (which in this calculation we will denote simply by $\tr$) are to be evaluated in the angular quantization scheme with the following boundary conditions:
\be\label{impozerofield1}
\phian\big(\eta =\log (\epsilon/x)\big) = \phian\big(\eta= \log (x/\ep)\big) =0.
\ee
We will evaluate the quantity
\be\label{ratioZZregAQ1}
\langle \T_{-\alpha}^\epsilon(0)\T_\alpha^\epsilon(x)\rangle_{\rm CFT} = \frac{ \tr \, [\exp( -2 \pi (\Han^{\mathrm{CFT}} - \ii \alpha \Qan^{\rm CFT}))]}{ \tr \, [\exp( -2 \pi \Han^{\mathrm{CFT}})]}
\ee
under \eqref{impozerofield1} and then evaluate the constant $\mathrm c_\alpha$ in accordance with \eqref{Balpha}.

Let $\Delta=-2\log(\ep/x)$, and define the shifted variable $\h\eta=\eta-\log(\ep/x)$. Consider the following Fourier decomposition:
\be\begin{aligned}
	\phian(\eta) &= \sum_{k\in\ZZ_+} \frac{1}{\sqrt{\pi k}}
	(c_k^{\phantom\dag}+d_k^\dag)
	\sin(\pi k\h\eta/\Delta)\\
	\Pian(\eta) &= \frac{1}{\ii\Delta} \sum_{k\in\ZZ_+} \sqrt{\pi k}
	(c_k^{\phantom\dag}-d_k^\dag)
	\sin(\pi k\h\eta/\Delta).
	\end{aligned}
\ee
This automatically satisfies the conditions \eqref{impozerofield1}. The inverse is
\be\begin{aligned}
	c_k &= \frac{1}\Delta \int_0^\Delta \dd\h\eta\,
	\sin(\pi k\h\eta/\Delta)
	\lt(\frac{\ii \Delta}{\sqrt{\pi k}}\Pian(\eta)
	+\sqrt{\pi k}\,\phian(\eta)\rt) \\
	d_k &= \frac{1}\Delta \int_0^\Delta \dd\h\eta\,
	\sin(\pi k\h\eta/\Delta)
	\lt(\frac{\ii \Delta}{\sqrt{\pi k}}\Piand(\eta)
	+\sqrt{\pi k}\,\phiand(\eta)\rt),
	\end{aligned}
\ee
and using the canonical commutation relations \eqref{canan}, we  have
\be\label{cancd}
	[c_k^{\phantom\dag},d_{l}^{\phantom\dag}]=[c^\dag_k,d^\dag_{l}]=0,\qquad
	[c_k^{\phantom\dag},c^\dag_{l}] = [d_k^{\phantom\dag},d_{l}^\dag] = \delta_{k,l}, \qquad k,l\in \mathbb Z_+.
\ee
The Hamiltonian \eqref{Han}, with $m=0$, takes the form
\be\begin{aligned}
	\Han^{\rm CFT} &= \int_0^{\Delta} \dd\h\eta\,\normord{\Piand(\eta)\Pian(\eta)
	+\p_\eta \phiand(\eta)\p_\eta\phian(\eta)} \\
	&= \sum_{k\in\ZZ_+} \nu_k^{\phantom\dag}
	\normord{c_k^\dag c_k^{\phantom\dag} + d_k^\dag d_k^{\phantom\dag}}
	\end{aligned}
\ee
where the energies are
\be
	\nu_k = \frac{\pi k}{\Delta}.
\ee
Therefore, the angular time-evolved fields are
\be\begin{aligned}
	\phian(\eta,\xi) &= \sum_{k\in\ZZ_+} \frac{1}{\sqrt{\pi k}}\,
	(c_k^{\phantom\dag}e^{-\nu_k\xi}+d_k^\dag e^{\nu_k\xi})
	\sin(\pi k\h\eta/\Delta)\\
	\Pian(\eta,\xi) &= \frac{1}{\ii\Delta} \sum_{k\in\ZZ_+} \sqrt{\pi k}\,
	(c_k^{\phantom\dag}e^{-\nu_k\xi}-d_k^\dag e^{\nu_k\xi})
	\sin(\pi k\h\eta/\Delta).
	\end{aligned}
\ee
We define the angular vacuum $|0\rangle_{\rm an}$ as usual by requiring the vanishing the fields $\phian(\eta,\xi)$, $\phiand(\eta,\xi)$ at negative infinite Euclidean time $\xi\to-\infty$, which imposes
\be
	c_k|0\rangle_{\rm an} = d_k|0\rangle_{\rm an} = 0,
	\qquad
	 k\in\ZZ_+.
\ee
The full angular Hilbert space is the Fock space of the relations \eqref{cancd} over this vacuum, and the normal-ordering is that where $c_k$'s and $d_k$'s are brought to the right. Finally, the conserved charge is given by
\be
\Qan^{\rm CFT} = \sum_{k\in\ZZ_+} (d^\dag_k d_k^{\phantom\dag} - c^\dag_k c_k^{\phantom\dag}).
\ee

Note that the Fock space is the infinite tensor product, over all discrete values of $k\in\ZZ_+$, of the tensor products of the two single-bosonic-mode spaces ${\rm span}\{(c_k^\dag)^l |0\rangle_{\rm an}:l\in\{0,1,2,\ldots\}\}$ and ${\rm span}\{(d_k^\dag)^l |0\rangle_{\rm an}:l\in\{0,1,2,\ldots\}\}$.  
Therefore, the trace in \eqref{ratioZZregAQ1} likewise factorizes and this quantity becomes
\be\label{eq_trproddnu12}\begin{aligned}
\langle \T_{-\alpha}^\epsilon(0)\T_\alpha^\epsilon(x)\rangle_{\rm CFT} &=  \frac{ \tr \lt[  \prod_{k=1}^\infty \exp \Bigl( -2 \pi  (\nu_k^{\phantom\dag} + \ii \alpha) c_k^\dag c_k^{\phantom\dag} - 2 \pi ( \nu_k^{\phantom\dag}  - \ii \alpha) d_k^\dag d_k^{\phantom\dag} \Bigr)\rt] }{ \tr\lt[  \prod_{k=1}^\infty \exp \Bigl( -2 \pi  \nu_k^{\phantom\dag}  c_k^\dag c_k^{\phantom\dag} - 2 \pi  \nu_k^{\phantom\dag}  d_k^\dag d_k^{\phantom\dag} \Bigr) \rt]}  \\
&= \prod_{k=1}^{\infty} \frac{ \tr_{\rm bos} \lt[ \exp \Bigl( -2 \pi  (\nu_k + \ii \alpha) \h n\Bigr)\rt]}{ \tr_{\rm bos} \lt[\exp \Bigl( -2 \pi  \nu_k  \h n\Bigr)\rt] } \,
 \frac{ \tr_{\rm bos} \lt[ \exp \Bigl( -2 \pi  (\nu_k - \ii \alpha) \h n\Bigr)\rt]}{ \tr_{\rm bos} \lt[\exp \Bigl( -2 \pi  \nu_k  \h n\Bigr) \rt]}  
\end{aligned}
\ee
where $\tr_{\rm bos}$ is the trace over a single bosonic mode, and $\h{{n}}$ is the number operator of this mode. Using $\tr_{\rm bos} q^{\h n} = (1-q)^{-1}$, we therefore obtain
\be\label{Zprod}
\langle \T_{-\alpha}^\epsilon(0)\T_\alpha^\epsilon(x)\rangle_{\rm CFT}= \prod_{k=1}^\infty \frac{(1-q^k)^2}{(1-t q^k) (1-t^{-1}q^k)}
\ee
with
\be
q=\exp(-2\pi^2/ \Delta)
, \qquad t=\exp(2\pi \ii \alpha).  
\ee
This expression depends on $\epsilon$ via $\Delta = -2\log(\ep/x)$. When  $\epsilon\rightarrow0$, i.e.  $q\rightarrow 1^-$, the asymptotic expansion of \eqref{Zprod} can be studied with the Mellin transform (see Appendix \ref{MTransfo}) and is
\be\label{Zexp}
\langle \T_{-\alpha}^\epsilon(0)\T_\alpha^\epsilon(x)\rangle_{\rm CFT}  \sim  \frac{4}{\pi} \, \sin (\pi \alpha)\, |\log(\ep/x)|\,\lt(\frac{\epsilon}x\rt)^{2h_\alpha}.
\ee
This indeed gives $\ell_2=1$ as claimed \eqref{ell12bos}, and reproduces \eqref{Zexpov}. 
Hence, taking the limit $\epsilon\rightarrow 0$ of this last expression with the scheme \eqref{Balpha}, one obtains the value of the renormalization constant \eqref{calphabos}.


\subsubsection{Regularized one-point function in the massive QFT}\label{s433}

We now consider \eqref{VEVtrace_eps1}. In order to evaluate the trace involved, we diagonalize simultaneously the (massive) Hamiltonian \eqref{Han} and the charge \eqref{Qan}, with the single boundary condition
\be\label{bdcondm}
	\phian(\eta = \log\ep) = 0.
\ee
Combining this with the fact that the mass barrier causes waves to reflect, the set of energies will be discrete. We denote by $\nu>0$ these energies.

Consider the decomposition in modes $a_\nu$ and $b_\nu$ as follows:
\be
	\phian(\eta,\xi) = \sum_{\nu>0} \sqrt{\frac2\nu}\lt(
	a_\nu e^{-\nu\xi} U_\nu(\eta) + b_\nu^\dag e^{\nu\xi}V_\nu(\eta)
	\rt),\quad
	\Pian(\eta,\xi) = -\ii\sum_{\nu>0} \sqrt{2\nu}\lt(
	a_\nu e^{-\nu\xi} U_\nu(\eta) - b_\nu^\dag e^{\nu\xi}V_\nu(\eta)
	\rt).
\ee
This implements the relation between $\Pian$ and $\phian$, along with the fact that $a_\nu^\dag$ and $b_\nu^\dag$ diagonalize the Hamiltonian with positive energies. Consistency then imposes that the partial waves $U_\nu(\eta)$ and $V_\nu(\eta)$ satisfy the equations of motion,
\be
\partial_\eta^2 U_\nu(\eta) - m^2 e^{2\eta} U_\nu(\eta) + \nu^2 U_\nu(\eta)=\partial_\eta^2 V_\nu(\eta) - m^2 e^{2\eta} V_\nu(\eta) + \nu^2 V_\nu(\eta)=0.
\ee
The general solution can be expressed in terms of modified Bessel functions, $C_1 I_{\ii\nu} (me^\eta) + C_2K_{\ii\nu}(me^\eta)$. The condition that partial waves vanish as $\eta\to\infty$ impose $C_1=0$, and since $K_{i\nu}(me^{\eta}) = K_{-i\nu}(me^{\eta})$, we may choose
\be
	U_\nu(\eta) = V_\nu(\eta)= \lt(\int_{\log\ep}^\infty\dd\eta\, K_{\ii\nu}(me^{\eta})^2\rt)^{-\frac12} K_{\ii\nu}(me^\eta).
\ee

Since the functions $U_\nu$ diagonalize a Hermitian differential operator, they must be orthogonal, and therefore
\be
	\int_{\log\ep}^\infty \dd\eta\,U_\nu(\eta)U_{\mu}(\eta) = \delta_{\nu,\mu},\;\qquad \nu,\mu>0.
\ee
We may then express the modes in terms of fields (here at $\xi=0$),
\be\begin{aligned}
	a_\nu &= \frac{1}{\sqrt{2}}\int \dd\eta\,U_\nu(\eta)\lt(
	\sqrt\nu\,\phian(\eta)+\frac{\ii}{\sqrt\nu} \Pian(\eta)\rt)\\
	b_\nu &= \frac{1}{\sqrt{2}}\int \dd\eta\,U_\nu(\eta)\lt(
	\sqrt\nu\,\phiand(\eta)+\frac{\ii}{\sqrt\nu} \Piand(\eta)\rt)
	\end{aligned}
\ee
and it is simple to see that
\be
	[a_\nu^{\phantom\dag},a_{\mu}^\dag] = [b_\nu^{\phantom \dag},b_{\mu}^\dag] = \delta_{\nu,\mu}
\ee
other commutators vanishing. We define the angular vacuum as usual, giving rise to $a_\nu|0\rangle_{\rm an}=b_\nu|0\rangle_{\rm an}=0$, and we construct the Fock space over this vacuum. Finally, the Hamiltonian and charge take the usual form,
\be
	\Han = \sum_{\nu>0} \nu(a_\nu^\dag a_\nu^{\phantom\dag} + 
	b_\nu^\dag b_\nu^{\phantom\dag}),\qquad
	\Qan = \sum_{\nu>0} ( 
	b_\nu^\dag b_\nu^{\phantom\dag} - a_\nu^\dag a_\nu^{\phantom\dag}).
\ee
By the same arguments as in \eqref{eq_trproddnu12}-\eqref{Zprod}, we must therefore evaluate
\be\label{toeval}
	\langle \T_\alpha^\epsilon \rangle = \frac{\tr_\ep\left(e^{-2\pi (H_{\rm an}-\ii\alpha Q_{\rm an})}\right)}{\tr_\ep\left(e^{-2\pi H_{\rm an}}\right)}
	=\prod_{\nu>0}^\infty \frac{(1-e^{-2\pi \nu})^2}{(1-t e^{-2\pi \nu}) (1-t^{-1}e^{-2\pi \nu})}
\ee
with $t=\exp(2\pi \ii \alpha)$.

The quantity \eqref{toeval} depends on the exact set of energies $\nu$. In order to determine this set, we consider the boundary condition \eqref{bdcondm}. The asymptotic form of $K_{\ii\nu}(me^\eta)$ for $\nu\neq0$ as $\eta\to-\infty$ contains both positive- and negative-frequency oscillating exponentials:
\be
K_{\ii\nu}(me^\eta) \sim \frac12 \Bigl( \Gamma(\ii\nu) \bigl( \frac{m}{2} \bigr)^{-\ii\nu} e^{-\ii\nu \eta} + \Gamma(-\ii\nu)   \bigl( \frac{m}{2} \bigr)^{\ii\nu} e^{\ii\nu \eta} \Bigr) , \qquad \nu\neq0, \; \eta\rightarrow -\infty
\ee
and correction terms are exponentially decaying. The set of energies is therefore determined by the condition
\be
	\Gamma(\ii\nu) \bigl( \frac{m}{2} \bigr)^{-\ii\nu} \ep^{-\ii\nu} + \Gamma(-\ii\nu)   \bigl( \frac{m}{2} \bigr)^{\ii\nu} \ep^{\ii\nu }=0
\ee
That is, $\nu=\nu_k$ with
\be
	\nu_k = \frac{(2\ii)^{-1}\log \mathrm{S}(\nu_k) - \pi k}{\log(m\ep/2)},\qquad k\in \ZZ_+
\ee
where the scattering phase off the mass barrier takes the form
\be
	\mathrm{S}(\nu) = e^{\ii\pi}\frac{\Gamma(\ii\nu)}{\Gamma(-\ii\nu)}.
\ee
Consider the variation
\be
	\nu_{k+1}-\nu_k = u\lt(1 - \frac{\mathrm{S}'(\nu_k)}{2\pi\ii \mathrm{S}(\nu_k)} u
	+ O(u^2)\rt),\qquad u = -\frac{\pi}{\log(m\ep/2)}
\ee
where $\mathrm{S}'(\nu)= \partial_\nu \mathrm{S}(\nu)$.
It is clear that to leading order in $u$, we have $\nu_k = ku$. Since the next correction is at one higher order in $u$, we expect that the use of $\nu_k=ku$ gives the correct divergent terms in $u^{-1}$ and $\log u$ in the asymptotic of $\log \langle \T_\alpha^\epsilon \rangle$. Therefore we may use \eqref{logZA} with $\Delta=-\log(m\ep/2)$ and we have
\be\label{1poQFTfirstorno1}
	\log \langle \T_\alpha^\epsilon \rangle \sim \alpha(1-\alpha)\log(m\ep/2)
	+\log(\log(2/(m\ep))) + \log \Bigl[ \frac2\pi \vert \sin ( \pi \alpha) \vert \Bigr] + O(1).
\ee

The exact remaining $O(1)$ term can be evaluated as follows. As $u\to0$, the variations $\nu_{k+1}-\nu_k$ tend to zero, and we may replace sums by integrals, i.e.
\be\label{repSumintoIntno1}
	\sum_{\nu>0} f(\nu) \; \rightarrow \;  \int_0^\infty \frac{\dd\nu}u \Bigl[
	1+ \frac{u}{2\pi\ii} \bigl( \p_\nu\log \mathrm S(\nu) \bigr) \, 
	\Bigr] \, f(\nu) + O(1)
\ee
whenever $f(\nu)$ is a regular enough function.  
A quick look at equation \eqref{toeval} reveals that the function $\log \langle \T_\alpha^\epsilon \rangle$ contains a singularity at $\nu=0$ and hence a direct use of prescription \eqref{repSumintoIntno1} is not valid. However, in the difference
\be\label{primediffnoQFTp1r}
	\sum_{\nu>0}\log\lt(\frac{(1-e^{-2\pi \nu})^2}{(1-t e^{-2\pi \nu}) (1-t^{-1}e^{-2\pi \nu})}\rt) - \sum_{\nu=ku,\,k\in\ZZ_+}\log\lt(\frac{(1-e^{-2\pi \nu})^2}{(1-t e^{-2\pi \nu}) (1-t^{-1}e^{-2\pi \nu})}\rt)
\ee
all terms near to this singularity are cancelled, up to terms of higher order in $u$, and therefore we may use  the formula \eqref{repSumintoIntno1} for evaluating this difference. Thus, the difference \eqref{primediffnoQFTp1r} becomes
\be
	 \int_0^\infty \frac{\dd\nu}{2\pi\ii} \, \bigl( \p_\nu\log \mathrm S(\nu) \bigr) \, \log \left(
	\frac{(1-e^{-2\pi \nu})^2}{(1-t e^{-2\pi \nu}) (1-t^{-1}e^{-2\pi \nu})} \right).
\ee
This integral may be evaluated using integration by part. We first obtain
\be\label{intdnuvers1pp1}
\frac12 \int_{-\infty}^\infty \dd \nu \, \int_0^\infty  \frac{\dd t}{t}
	e^{-t}\lt(\nu - \frac{\sin \nu t}{1-e^{-t}}\rt) \,
	\bigl(\coth \pi(\nu+\ii\alpha) + \coth \pi(\nu-\ii\alpha) - 2\coth \pi\nu\bigr)
\ee
where we have used the integral representation
\be
	\log \mathrm S(\nu) = 2\ii\int_0^\infty \frac{\dd t}{t}
	e^{-t}\lt(\nu - \frac{\sin \nu t}{1-e^{-t}}\rt).
\ee
The integration over $\nu$ in \eqref{intdnuvers1pp1} can be performed using contour deformation, with the use of the identity
\be
	\int_{-\infty-\ii0}^{\infty-\ii0}
	\frac{\dd\nu}2\,e^{q \nu}(
	\coth \pi(\nu+\ii\alpha) + \coth \pi(\nu-\ii\alpha) - 2\coth \pi\nu
	)
	=
	\frac{e^{-\ii q  /2}-\cos q(1/2-\alpha)}{\sin(q/2)}
\ee
for $\alpha\in[0,1]$, and the result is
\be
	\int_0^\infty \frac{\dd t}{t} e^{-t}
	\lt(     \frac{2\sinh \bigl( \frac{t\alpha} 2 \bigr) \,\sinh \bigl( \frac{t(1-\alpha)}2\bigr)}{(1-e^{-t})\, \sinh \bigl( \frac t2 \bigr)}  -\alpha(1-\alpha)  \rt).
\ee
Therefore, taking into account the sum on the right in the difference \eqref{primediffnoQFTp1r}, from which the contribution is given by \eqref{1poQFTfirstorno1}, we obtain
\be\label{Zm}
	\langle \T_\alpha^\epsilon \rangle = 
	\exp\lt[
	\int_0^\infty \frac{\dd t}{t} e^{-t}
	\lt( \frac{2\sinh \bigl( \frac{t\alpha} 2 \bigr) \,\sinh \bigl( \frac{t(1-\alpha)}2\bigr)}{(1-e^{-t})\, \sinh \bigl( \frac t2 \bigr)} -h_\alpha\rt)
	\rt]\,
	\frac2\pi\, \sin( \pi \alpha)\,|\log(m\ep/2)|\,\lt(\frac{m\ep}2\rt)^{h_\alpha}
\ee
with (recall that) $h_\alpha= \alpha(1-\alpha)$,  the scaling dimension of $\mathcal T_\alpha$. This indeed gives $\ell_1=1$ as claimed \eqref{ell12bos}.

\subsubsection{Logarithmic structure, normalization constant, VEV of $\T_\alpha$} 
Observe that, when comparing \eqref{Zexp} and \eqref{Zm}, the general concepts explained in Subsection \ref{ssectreg} are indeed correct: both the CFT regularized two-point function \eqref{Zexp} and the massive QFT regularized one-point function \eqref{Zm} present the same logarithmic behavior in the UV cutoff $\ep$, with the logarithmic powers chosen as per \eqref{ell12bos}. 
The logarithm indicates the presence of non-compactness, and, as explained, the renormalization necessary in order to obtain a universal expression is that where this logarithm is divided out: we see that indeed the choice \eqref{ell12bos} is the only possible choice that gives finite results.  Thus, using the renormalization constant $\mathrm c_\alpha$ obtained in the previous section from the CFT two-point function and the prescription \eqref{logmass} with $\langle \T_\alpha^\epsilon\rangle$ in \eqref{Zm}, the expression \eqref{VEVboson} for the VEV is obtained. \\

\noindent
\emph{Remark on the boundary conditions \eqref{impozerofield1} and \eqref{bdcondm}.}  We have defined the regularized twist field using the Dirichlet boundary conditions, \eqref{impozerofield1} and \eqref{bdcondm}, and from this, we have obtained the renormalized twist field correlation functions, see \eqref{logmass}, \eqref{logCFT} and \eqref{logmassplus}. Another natural choice of boundary conditions is that of the Neumann boundary conditions, $\partial_\eta\phian(\eta=\log(\epsilon))=0$. In this case, a computation in angular quantization can also be done and is similar to the one presented here. Of course, in this case, because of the unbounded zero mode, the regularized twist field is {\em not} well defined in CFT.  We show in Appendix \ref{appneu} that, with Neumann boundary conditions and under the special CFT normalization according to which the infinite zero-mode trace is made to diverge proportionally with $|\log\epsilon|$, there is {\em no logarithmic renormalization}, i.e.~$\ell_1=\ell_2=0$, and thus no logarithmic factors appear in \eqref{Zexpov} and \eqref{Zexpovmass}; and the VEV of the renormalized twist field is again exactly given by \eqref{VEVboson}. This indicates that the {\em same} renormalized twist field is obtained, and in particular \eqref{ttov} should still hold.


\section{Application to universal entanglement saturation: branch-point twist fields}\label{sectE}

In this section we make the connection between the entanglement entropy (EE) and the results of the preceding sections.  We begin by recalling the construction of the EE.

Let $\mathcal H$ be the Hilbert space of a one-dimensional quantum system and denote by $A$ a given subregion of the system.  The complement of $A$ is denoted by $B$. Typically, one can think of $\mathcal H$ as representing a one-dimensional quantum spin chain, and $A$ and $B$ being two (disjoint) complementary intervals on the spin-chain.  
Note that later, we will be interested in the scaling limit of the spin chain, where it can be  described in terms of a QFT (or a CFT) so $A$ is a finite interval in space. We also take the system to be infinite in size.   
Let  $\vert \psi\rangle$ be the state representing the ground state of the quantum system, and denote its density matrix by $\rho=|\psi\rangle\langle\psi|$.  Since $\mathcal H =   {\mathcal H}_A\otimes {\mathcal H}_B$, where ${\mathcal H}_A$ (resp. ${\mathcal H}_B$) corresponds to the Hilbert space associated to the degrees of freedom of subsystem $A$ (resp. $B$), we define the reduced density  matrix $\rho_A$ by tracing out $\mathcal H_B$, i.e. $\rho_A = \tr_{{\mathcal H}_B} \rho$.  
  The EE $S^{(n)}$ is a bi-partite entanglement measure, between subregions $A$ and $B$,  defined as the R\'enyi entropy:
\be\label{EEno12}
	S^{(n)} = \frac1{1-n}   \log \tr_{{\mathcal H}_A} (\rho_A^{n}).
\ee
with $n>0$.    The limit $n=1$ is referred to as the von Neumann entropy,
\be
S^{(1)} = \lim_{n\to 1}S^{(n)} = -\tr_{{\mathcal H}_A} (\rho_A\log\rho_A).
\ee
The quantity $S^{(n)}$ provides a good measure of entanglement for pure states.  
The logarithmic negativity may be defined in somewhat related ways and provides a good measure for mixed states (see \cite{Plenio_LN05,CalabCardy12_LN1, Plenio05_revEE} for the details).

The main results of this section are as follows. Let $\xi$ be the correlation length and $L$ the length of the connected region $A$.  Let us also denote by $S^{(n)}_{\xi;L}$ the R\'enyi EE associated to a given correlation length $\xi$ and region $A$ of length $L$.
Then the following difference has a universal expansion, involving the VEV of the branch-point twist field $\T$ (see below) via $\mathsf V_\T= m^{-h_n}\langle\T\rangle$:
\be\label{dblimlog}
	\lim_{b\to\infty}
	\lt(S^{(n)}_{a;\,b} - S^{(n)}_{b;\,a}\rt)
	= \frac{2\ell_1^{(n)}-\ell_2^{(n)}}{1-n}\log(\log a)
	+\frac2{1-n}\log(\mathsf V_\T) + o(1),\qquad \; a\to\infty.
\ee
This generalizes the formula when no logarithmic renormalization is present \cite{CD09}, i.e.
\be\label{dblim}
	\lim_{a\to\infty}\lt(\lim_{b\to\infty}
	\lt(S^{(n)}_{a;\,b} - S^{(n)}_{b;\,a}\rt)\rt) = \frac{2}{1-n} \log(\mathsf V_\T).
\ee
In the case of the one-dimensional Klein-Gordon model (the massive real free boson), we have:
\be\label{VTKG}
	\ell_1^{(n)} = \ell_2^{(n)} = \frac{n-1}2,\quad
	\mathsf V_{\T,\,{\rm KG}} = 2^{\frac{(1-n)(1+4n)}{12n}} n^{\frac14} \pi^{\frac{1-n}4} 
	\exp \Bigl[ \int_0^\infty \frac{\dd t}{2t}  e^{-t}  \Bigl( \frac{n \coth\bigl( \frac t2\bigr)  -\coth\bigl( \frac t{2n}\bigr) }{(1-e^{-t})} - \frac16\lt(n-\frac1n\rt)\Bigr) \Bigr].
\ee

In particular, let us consider for simplicity the von Neumann EE ($n=1$). In this case, the full logarithmic structure and saturation in the massive real free boson is as follows. We find that in the limit $L\to\infty$, the asymptotic $\xi\to\infty$ is
\be\label{S1fb}
S^{(1),{\rm KG}}_{\xi;\infty}= \frac{1}{3} \log(2\xi/\mathrm A_1)-\log (\log\xi)  + \frac{ \log(2\pi)-1}2 +\int_0^\infty \frac{\dd t}{t} e^{-t} \Bigl( \frac{\sinh(t) - t}{(\cosh(t) -1)(e^{-t}-1)} + \frac{1}{3} \Bigr) + o(1)
\ee
where $\rm A_1$ is a non-universal constant (see below). In the reversed limit order, we obtain
\be\label{S1fb2}
	S^{(1),{\rm KG}}_{\infty;\, L} = \frac{1}3 \log(L/\mathrm A_1)
	-\frac12\log(\log(L))
	+ o(1)
\ee
where the limit $\xi\to\infty$ is first evaluated, then followed by the asymptotic $L\to\infty$.  Furthermore, in the scaling limit $L,\,\xi\to\infty$ with $L/\xi$ fixed, the asymptotic is
\be\label{S1fb3}
	S^{(1),{\rm KG}}_{L;\xi} = \frac{1}3 \log(\xi/\mathrm A_1)
	-\log (\log\xi)+F^{(1)}(L/\xi) + o(1)
\ee
where $F^{(n)}(mx) = \log\lt(m^{-2h_n}\langle \T(x)\T(0)\rangle\rt)$ is a dimensionless universal scaling function, whose asymptotic $L/\xi\to\infty$ is a constant reproducing (at $n=1$)  \eqref{S1fb}, and whose asymptotic $L/\xi\to 0$ is (at $n=1$)
\be\label{S1fb4}
	F^{(1)}(L/\xi) = \frac13 \log (L/\xi)+\frac12\log (\log(L/\xi))+o(1).
\ee
Note that in all these expressions, the same constant $\mathrm A_1$ appears.

\subsection{Branch-point twist fields}

Branch-point twist fields are defined as follows. Consider a model of QFT; it has a Hilbert space $\mathcal H$, and may be represented by an action $S[\varphi]$, or more formally, by a set of local fields $\mathrm V = \{\Or\}$ and an operator algebra amongst these,
\be
	\Or(0) \Or'(x) = \sum_{\Or''} C_{\Or,\Or'}^{\Or''}(x) \Or''(0).
\ee
Now consider $n$ independent copies of this model. This is a new model of QFT, whose Hilbert space is the $n^{\rm th}$ tensor product of the Hilbert space of the original model, ${\mathcal H}_n = {\mathcal H}^{\otimes n}$. The $n$-copy model contains the set of fields formed by the $n$-fold tensor product $\mathrm V_n = \mathrm V \otimes \cdots\otimes \mathrm V$. The interactions factorize amongst the copies, and accordingly, denoting $\underline{\Or}=(\Or_1,\ldots,\Or_n)\in \mathrm V_n$, the operator algebra is
\be
	\underline{\Or}(0)\, \underline{\Or}'(x) =
	\sum_{\{\Or''\}} \lt(\prod_{j=1}^n C_{\Or_j^{\phantom \prime}, \Or_j'}^{\Or_j''}(x)\rt)
	\underline{\Or}''(0).
\ee
If an action exists with path-integral field $\varphi$, then denoting $\varphi_j = ({\bf 1},\ldots,{\bf 1},\varphi,{\bf 1},\ldots,{\bf 1})$ where the nontrivial factor $\varphi$ is at position $j$, the action of the $n$-copy model is the function of $\{\varphi\} = \{\varphi_j:j=1,\ldots,n\}$ given by
\be
	S[\{\varphi\}] = \sum_{j=1}^n S[\varphi_j].
\ee

The $n$-copy model has a natural internal permutation symmetry. In particular, it has a $\ZZ_n$ symmetry subgroup, that under cyclic permutation of the copies generated by $\sigma$:
\be
	\sigma\cdot (\Or_1,\ldots,\Or_n) = 
	(\Or_n,\Or_1,\ldots,\Or_{n-1}).
\ee
This is indeed a symmetry of the action, since $S[\{\sigma\cdot\varphi\}] = S[\{\varphi\}]$, and it preserves the operator algebra. 
According to the concept presented in Section \ref{sectoverview}, one may therefore construct twist fields associated to elements of this internal symmetry. Denote the twist field associated to $\sigma$ by $\T=\T_{\sigma}$, and that associated to the inverse element by $\tT = \T_{\sigma^{-1}}$ (sometimes called the anti-twist field).  Note that in the quantization on the line, we have $\T^\dag = \tT$. These were first studied in general QFT in \cite{CCD1}, where they were named {\em  branch-point twist fields}. Such fields were also studied widely in the context of orbifold CFT and in relation with partition functions on Riemann surfaces. The scaling dimension of primary branch-point twist fields in the $n$-copy model is \cite{knizhnik87}
\be\label{hn}
	h_n = \frac{c}{12} \Bigl(n-\frac1n\Bigr)
\ee
where $c$ is the central charge of the corresponding CFT.   In massive integrable QFT, form factors of branch-point twist fields were first studied in \cite{CCD1}.

We note that it is always possible to organize $\mathrm V_n$ into eigenspaces of the symmetry $\sigma$. Given $\underline{\Or}\in \mathrm V_n$, we may construct its Fourier transform 
\be\label{FouTransfoOrno3}
	\hat{\underline{\Or}}^{(k)} = \sum_{j=1}^n e^{-2\pi \ii j k/n}\, \sigma^j\cdot
	\underline\Or,\qquad k=0,\ldots,n-1.
\ee
Such fields have the property that $\sigma\cdot\hat{\underline{\Or}}^{(k)} = e^{2\pi \ii k/n} \hat{\underline{\Or}}^{(k)} $, and likewise, we have the decomposition
\be\label{decompoTV1}
 \mathrm V_n=\oplus_{k=0}^{n-1} \mathrm V_n^{(k)}, \qquad  \mathrm V_n^{(k)}=\{ \hat{\underline\Or}\in \mathrm V_n \; | \; \sigma\cdot \hat{\underline\Or} = e^{2\pi \ii k/n} \hat{\underline\Or}\}.
 \ee
On this decomposition, we therefore find, following \eqref{exch}, that
\be\label{exchbp}
	\T(x) \hat{\underline\Or}(y) = \left\{\begin{array}{ll} \hat{\underline\Or}(y) \T(x) & (x_1>y_1) \\
	e^{2\pi \ii k/n}
	\hat{\underline\Or}(y) \T(x) & (x_1<y_1) \end{array}\right.,
	\qquad
	\hat{\underline\Or} \in V_n^{(k)},
	\qquad x_2=y_2.
\ee
In particular, the space $\mathrm V_n^{(0)}$ forms an operator subalgebra, on which $\T$ is local in the strong sense (it commutes with all its elements at space-like distances). This is the basis for the orbifold construction of modules of vertex operator algebras \cite{DongMason1}, see also \cite{Dixon, Verlinde89}.

\subsection{Universal EE saturation}\label{ssectunisat}

The basic formula relating branch-point twist fields to the EE of extended quantum systems associates a branch-point twist field to every boundary point of the region $A$, and relates $\tr_{\mathcal H_A}(\rho_A^n)$ to the average of the product of such fields. The correct identification, near critical point as described by QFT, involved UV-regularized branch-point twist fields. Using the notation $\T^\ep$, $\tT^\ep$ of Subsection \ref{ssectreg} for the UV-regularized version of the field, in the single-interval case $A=[0,x]$ we have
\be
	\tr_{\mathcal H_A}(\rho_A^n) = C_n\,
	\langle \T^\ep(0) \tT^\ep(x)\rangle
\ee
where on the right-hand side, the average is in the ground state of the $n$-copy model. The proportionality constant $C_n$ is not universal. This equation allows us to express QFT quantities (on the right-hand side) in terms of the corresponding quantities in the microscopic model (on the left-hand side, say a spin chain). Observe that the EE \eqref{EEno12} is directly related to the correlation functions of {\em regularized branch-point twist field}, instead of those of the usual renormalized field.
Let us denote the
 \emph{lattice spacing} by $\chi$, the correlation length by $\xi=(m\chi)^{-1}$, and the  length of the interval on the chain by $L=x/\chi$.  In the limit $\epsilon\rightarrow 0$, the lattice spacing $\chi$ and the UV regularization parameter $\epsilon$ are related to each other via (another) non-universal constant: $\ep = B_n \chi$. The universal information about the EE is found in the universal scaling regime $\ep\to0$ (that is, $L,\xi\to\infty$ with fixed ratio $L/\xi=mx$).

Let us first repeat the usual argument extracting the divergency structure of the EE, without logarithmic renormalization \cite{Holzhey94,VLRK_03,CalabCardy_EE04,CCD1}. Thus assume that the universal branch-point twist field is obtained by a usual power-law renormalization, such as in \eqref{nolog}, and that this renormalization is the same in the massive model and in the CFT (that is, at finite but large, or infinite, correlation length):
\be
	\T = \lim_{\ep\to0} {\mathrm c}_n \ep^{-h_n} \T^\ep.
\ee
As usual, the real constants $\mathrm c_n$ are chosen such that $\langle \T(0)\tT(x)\rangle_{\rm CFT} =  |x|^{-2h_n}$. Then one finds in the scaling limit
\be\label{Snmass}
	S^{(n)} \sim -\frac{c}6 \lt(1+\frac1n\rt)\log(\mathrm A_n \chi)
	+\frac1{1-n}\log\langle \T(0)\T(x)\rangle + o(1) \qquad(\ep\to0)
\ee
where $\mathrm A_n = (C_n \mathrm c_n^{-2})^{1/(2h_n)}B_n$ is non-universal, and where the correction terms $o(1)$ vanish in the scaling limit $\ep\to0$. This formula indicates that, in the scaling limit, the EE diverges logarithmically. The universal information is in the $O(1)$ part of this asymptotics, and is a function of the dimensionless combination $mx$. It is universal up to the addition of a constant. If the correlation length is infinite (the model is critical), then the universal regime is described by a CFT and we have
\be
	S^{(n)} \sim \frac{c}6 \lt(1+\frac1n\rt)\log(L/\mathrm A_n )
	+o(1), \; \qquad(1\ll L \ll \xi).
\ee
That is, the EE diverges logarithmically with the length of the interval. Note that the short distance limit $mx\to0$ of the scaling form \eqref{Snmass} of $S^{(n)}$ is the same as $S_{\rm CFT}^{(n)}$, both representing the regime $1\ll L \ll \xi$. In a given model, one may therefore identify $\mathrm A_n$ by evaluating the $O(1)$ part of the critical EE $S^{(n)}_{\rm CFT}$, and once identified, QFT predicts a universal $O(1)$ part in the full scaling limit as per \eqref{Snmass}. 
In particular, the VEV of the branch-point twist field $\langle\T\rangle$ is associated to the {\em universal saturation} of the EE,
\be\label{eqEEsaturation_no123}
	S^{(n)} \sim \frac{c}6 \lt(1+\frac1n\rt)\log(\xi/\mathrm A_n)
	+\frac2{1-n}\log (\mathsf V_\T) + o(1), \;
	\qquad(1\ll \xi \ll L)
\ee
where we recall that $\mathsf V_\T= m^{-h_n}\langle\T\rangle$ is a pure number. That is, the EE saturation, obtained in the limit $L\to\infty$, diverges logarithmically with the correlation length $\xi$, and subtracting this divergency along with the appropriate constant $-\frac{c}6 \lt(1+\frac1n\rt)\log (\mathrm A_n)$ found at criticality, one obtains a universal constant, related to and predicted by the connection problem of QFT. One can also express this as the double limit \eqref{dblim}. Universal saturations of the logarithmic negativity in various configurations are also related to VEVs of branch-point twist fields, see \cite{BCD1, BC16_BranchP}.

Clearly, logarithmic scaling \eqref{logmass}, \eqref{logCFT}, \eqref{logmassplus} will modify the above arguments. Let us analyze the consequences on the definition of the universal EE saturation.

Taking first the large-distance limit as in \eqref{logmass}, there is clustering and we use
\be
	\langle \T\rangle = \mathrm c_n
	\lim_{\ep\to0} \frac{\ep^{-h_n}}{|\log\ep|^{\ell_{1}^{(n)}}}
	\langle \T^\ep\rangle
\ee
in order to obtain
\be\label{Snlogmass}
	S^{(n)} \sim \frac{c}6 \lt(1+\frac1n\rt)\log(\xi/\mathrm A_n)
	+\frac{2\ell_{1}^{(n)}}{1-n}\log (\log\xi)
	+\frac2{(1-n)}\log(\mathsf V_\T) + o(1),
\ee
for $L\to\infty$, then $\xi\to\infty$.  
The limit $L\to\infty$ is finite, and the $\xi\to\infty$ behavior presents both logarithmic and double-logarithmic divergencies. On the other hand, taking first the critical limit as in \eqref{logCFT}, we use
\be
	\langle \T(0) \tT(x)\rangle_{\rm CFT}
	= \mathrm c_n^2 \,\lim_{\ep\to0} \frac{\ep^{-2h_n}}{|\log\ep|^{\ell_2^{(n)}}}
	\langle\T^\ep(0)\tT^\ep(x)\rangle_{\rm CFT}
\ee
and we obtain
\be\label{SnlogCFT}
	S^{(n)} \sim \frac{c}6 \lt(1+\frac1n\rt)\log(L/\mathrm A_n)
	+\frac{\ell_{2}^{(n)}}{1-n}\log(\log(L))
	+ o(1),
\ee
for $\xi\to\infty$, then $ L\to\infty$.
The limit $\xi\to\infty$ is finite, and the $L\to\infty$ behavior presents both logarithmic and double-logarithmic divergencies.  The combination of \eqref{Snlogmass} and \eqref{SnlogCFT} provides a universal meaning of the VEV of twist field in terms of EE saturation. In particular, the limit in \eqref{dblimlog} has the universal expansion displayed.

If $2\ell_{1}^{(n)}\neq\ell_{2}^{(n)}$, then the double-logarithmic divergencies in correlation length \eqref{Snlogmass} and in interval length \eqref{SnlogCFT} are different.  Thus, it does not make sense to talk about the regime $1\ll L \ll \xi$.
Indeed using the form \eqref{logmassplus}, the universal scaling limit $L,\xi\to\infty$, keeping a fixed ratio $L/\xi$, is
\be\label{Snmasslog}
	S^{(n)} \sim -\frac{c}6 \lt(1+\frac1n\rt)\log(\mathrm A_n\chi)
	+ \frac{2\ell_1^{(n)}}{1-n}\log(\log \xi)
	+\frac1{1-n}\log\langle \T(0)\T(x)\rangle + o(1) \qquad(\ep\to0).
\ee
and the limit $L/\xi\to0$ of this is
\be\label{Snlogmassplus}\begin{aligned}
	S^{(n)} \sim \frac{c}6 \lt(1+\frac1n\rt)\log(L/\mathrm A_n)
	+\frac{2\ell_{1}^{(n)}}{1-n}\log (\log\xi)
	&+\frac{\ell_2^{(n)}-2\ell_{1}^{(n)}}{1-n}\log (\log(L/\xi))+o(1), \\
	&
	(L,\xi\to\infty,\ L/\xi\ \mbox{fixed};\ \mbox{then}\ L/\xi\to0),
	\end{aligned}
\ee
which is different from the $L\to\infty$ asymptotics of the limit $\xi\to\infty$, see \eqref{SnlogCFT}.

Note that because of logarithmic partners, correction terms on both asymptotics above should be of the order of $1/|\log (m\ep)|$, which are not very small and may make any numerical measurement of universal entanglement saturations rather difficult.

\subsection{The massive free boson}

We now evaluate the value of the VEV of the branch-point twist field for the massive free boson.  We will use the results obtained from the previous sections on the complex Klein-Gordon model, by relating the branch-point twist field to a product of $\grp U(1)$ twist fields.

Consider the complex Klein-Gordon model. The field $\phi(x)$ becomes in the replicated model a set of $n$ independent complex fields $\phi_j(x)$, with $j=1,\ldots,n$.  Each of these fields satisfy the same equation of motion thanks to the factorization of the operator algebra.  Consider the Fourier transform, as in \eqref{FouTransfoOrno3}:
\be
	\hat\phi^{(k)} = \sum_{j=1}^n e^{-2\pi \ii jk/n}\phi_j  \; \in \; \mathrm V_n^{(k)}, \; \quad \qquad k=0,\ldots,n-1
\ee
with the property $\sigma\cdot\hat\phi^{(k)} = e^{2\pi \ii k/n}\hat\phi^{(k)}$. Since the fields $\{{\bf 1},\phi_j:j=1,\ldots,n\}$ generate the whole space $\mathrm V_n$ under the operator algebra, so do the Fourier transformed fields $\{{\bf 1},\hat\phi^{(k)}:k=0,\ldots,n-1\}$. We may consider the subspace of fields $\hat{\mathrm V}_{n,k}$ generated by ${\bf 1}$ and $\hat\phi_k$, and since $\sigma$ acts diagonally on both, we have
\be\label{inv}
	\sigma\cdot \hat{\mathrm V}_{n,k} = \hat{\mathrm V}_{n,k}.
\ee
Note that $\hat{\mathrm V}_{n,k}$ is different from $\mathrm V_n^{(k)}$, in particular in $\hat{\mathrm V}_{n,k}$ all integer powers of $e^{2\pi \ii k/n}$ occur as $\sigma$-eigenvalues. The particularity of a free model is that there is a factorization
\be\label{Vdec}
\mathrm V_n = \hat{\mathrm V}_{n,0}\otimes \cdots \otimes \hat{\mathrm V}_{n,n-1}
\ee
and that {\em the operator algebra factorizes} on the tensor factors of \eqref{Vdec}. Thanks to \eqref{inv}, as a consequence $\sigma$ acts as a symmetry on each operator algebra $\hat{\mathrm V}_{n,k}$; we will denote this action by $\sigma_k$ (acting trivially on $\hat{\mathrm V}_{n,k'}$ for $k'\neq k$):
\be
	\sigma_k = {\rm id}\otimes \cdots\otimes {\rm id}\otimes \sigma|_{\hat{\mathrm V}_{n,k}}\otimes {\rm id}\otimes \cdots\otimes {\rm id}.
\ee
This means that there is an enhanced symmetry, as any product of $\sigma_k$s is a symmetry of the full operator algebra. This also means that we may decompose the branch-point twist fields as
\be\label{twistprod}
	\T =  \prod_{k=0}^{n-1} \T_{\sigma_k}, \qquad \quad \T_{\sigma_0} = {\rm id}.
\ee
 This holds both for the renormalized twist field and the regularized one, as the boundary condition $\phi_j(x=\ep e^{\ii\theta})=0$ implies the same boundary condition on each $\hat\phi^{(k)}$. From the action perspective, we simply have
\be
	S[\{\phi\}] = \sum_{k} S[\hat\phi^{(k)}]
\ee
so that the operator algebra $\hat{\mathrm V}_{n,k}$ is a single-copy complex Klein-Gordon model based on $\hat\phi^{(k)}$ and $\sigma_k$ is the element $e^{2\pi \ii k g/n}$ of the $\grp U(1)$ symmetry group of this single copy. Therefore, we have
\be\label{TTwfactoksn2}
	\T = \prod_{k=1}^{n-1} \T_{k/n},\quad
\ee
and equivalently for $\T^\ep$. The field $\T$ ($\T^\ep$) is the renormalized (regularized) branch-point twist field in the $n$-copy complex Klein-Gordon model $\mathrm V_n$, and the field $\T_{k/n}$ ($\T_{k/n}^\ep$) is a renormalized (regularized) $\grp U(1)$ twist field in the single-copy Klein-Gordon model $\hat{\mathrm V}_{n,k}$. This provides the connection between branch-point twist fields and $\grp U(1)$ twist fields in the complex Klein-Gordon theory.  
The factorization property \eqref{TTwfactoksn2} implies the factorization of correlation functions, for instance
\be\label{ident}
	\langle \T^\ep(0)\tT^\ep(x)\rangle = \prod_{k=1}^{n-1}
	\langle \T_{k/n}^\ep(0)\tilde{\T}_{k/n}^\ep(x)\rangle
\ee
(and the same relations hold for the renormalized fields). 

Recall that the scaling dimension for the $\grp U(1)$ twist field $\T_\alpha$ corresponding to the Klein-Gordon model is $\alpha(1-\alpha)$.  Using the factorization property, the scaling dimension of $\T$ is thus
\be
  h_n = \sum_{k=1}^{n-1} \frac kn\lt(1-\frac kn\rt)= \frac{1}{6} \lt( n - \frac1n\rt) 
\ee
which corresponds to the correct scaling dimension \eqref{hn} with $c=2$, the central charge associated to the model. 

The relation between correlation functions for the branch-point twist field and correlation functions for the $\grp U(1)$ twist field, see \eqref{ident}, allows us to use the results of Subsection \ref{ssectQuanAnKG} in order to fix all quantities involved in the general results of Subsection \ref{ssectunisat}.  
In particular, one finds
\be
	\ell_1^{(n)} = \ell_2^{(n)}= n-1
\ee
for the complex Klein-Gordon theory.   
From the VEV \eqref{VEVboson} and using the identities 
\be
\prod_{k=1}^{n-1} \sin(\pi k/n) = 2^{1-n} \, n, \qquad \quad \sum_{k=1}^{n-1} \frac{2 \sinh\bigl( \frac{tk}{2n}\bigr) \sinh\bigl(\frac{t}{2}(1- \frac{k}{n}) \bigr)}{\sinh\bigl(\frac{t}{2}\bigr)} = n \coth\Bigl( \frac t2\Big)-  \coth\Bigl( \frac{t}{2n}\Bigr),
\ee
we have
\be
	\mathsf V_\T= m^{-h_n}\langle\T\rangle = 2^{-h_n +(1-n)/2} n^{1/2} \pi^{(1-n)/2} 
	\exp \Bigl[ \int_0^\infty \frac{\dd t}{t}  e^{-t}  \Bigl( \frac{n \coth\bigl( \frac t2\bigr)  -\coth\bigl( \frac t{2n}\bigr) }{(1-e^{-t})} - h_n\Bigr) \Bigr].
\ee
When $n\rightarrow 1$, we have the value $\lim_{n\rightarrow 1} \mathsf V_\T=1$ since the branch-point twist field becomes the identity field.  If we consider $n\in \mathbb R$ then we find
\be
\lim_{n\rightarrow 1} \frac{\partial}{\partial n} \mathsf V_\T = - \frac{c}{6} \log (2) + \frac12 \Bigl( 1 - \log(2\pi) \Bigr) - \int_0^\infty \frac{\dd t}{t} e^{-t} \Bigl( \frac{\sinh(t) - t}{(\cosh(t) -1)(e^{-t}-1)} + \frac{c}{6} \Bigr)
\ee
where $c=2$.   The universal saturation to the von Neuman EE is, using \eqref{Snlogmass},
\be
S^{(1)}\sim \frac{c}{3} \log(\xi/\mathrm A_1)-2\log (\log\xi) + \frac{c}{3} \log (2) -  \Bigl( 1 - \log(2\pi) \Bigr) +2 \int_0^\infty \frac{\dd t}{t} e^{-t} \Bigl( \frac{\sinh(t) - t}{(\cosh(t) -1)(e^{-t}-1)} + \frac{c}{6} \Bigr).
\ee

We end this section by mentioning that we have obtained the VEV for the complex Klein-Gordon theory.  
Using the fact that the complex Klein-Gordon field theory is the direct product of two real Klein-Gordon models, and the fact that the symmetry $\sigma$ factorizes as a product of independent permutation symmetry on each factor, it is straightforward to obtain the VEV of the real massive Klein-Gordon from the correspondance 
\be\label{ident2}
	\langle \T\rangle_{\mbox{\small complex KG}} = 
	\langle \T\rangle_{\mbox{\small real KG}}^2.
\ee

\section{Conclusion}

In this paper, we have undertaken the study of vacuum expectation values (VEVs) for twist fields associated to an internal symmetries of QFT models. The main results of this study are as follows.

First, we presented a new differential equation that permits the exact evaluation of VEVs of twist fields associated to elements of a Lie group in the image of the exponential map, see \eqref{Eq_diff_QFTno3}. The equation gives the VEV under the correct CFT normalization, and thus solves the QFT connection problem for this class of fields. It involves the two-point function between the twist field and the conserved Noether current related to the symmetry, which can be evaluated using a form-factor expansion.  This expansion is typically infinite, but becomes exact at two-particle order for free theory. As illustrations, we considered two simple cases, namely the $\grp U(1)$ twist fields in the Dirac fermion theory and in the complex Klein-Gordon boson theory. 

Second, we described the logarithmic renormalization of twist fields in models with non-compact target space, and the corresponding short-distance logarithmic behaviors of massive two-point functions. A proper consideration of logarithmic renormalization is essential in order to give the VEV a universal meaning. Using -- and clarifying -- the techniques of angular quantization, we studied the example of the  $\grp U(1)$ twist field in the complex Klein-Gordon model, extracting the exact logarithmic divergences both in the massless and massive regimes, see \eqref{Zexpov}, \eqref{Zexpovmass} and \eqref{ttov}. Using these techniques we also obtained its VEV in an alternative form \eqref{VEVboson}, confirming (numerically) agreement with the expression obtained by the general formula \eqref{Eq_diff_QFTno3}.

Finally, we applied these concepts to the study of the entanglement entropy, via the use of the branch-point twist field. In this context, the VEV of the branch twist field is related to the universal saturation of the EE. We explained how the logarithmic structure of the renormalization changes the way the universal saturation is extracted, which must take into account extra $\log \log$-type behaviors, see \eqref{dblimlog}. We applied these to the EE of the massive real one-dimensional Klein-Gordon boson (using the fact that the complex boson is formed of two independent copies of the real boson), see \eqref{VTKG}, and we obtained the logarithmic structure of the EE in various regimes, see \eqref{S1fb}, \eqref{S1fb2} and \eqref{S1fb3}.\\

We make few remarks.
\begin{enumerate}
\item As already noticed in \cite{CCD1}, an alternative method, within angular quantization,  to that presented here for the evaluation of CFT two-point functions $\langle \T(0)\tT(x)\rangle_{\rm CFT}$ and QFT one-point function $\langle \T(0)\rangle$ of branch-point twist fields is to account for the angle of $2\pi n$, instead of $2\pi$, around the origin as follows:
\[
	\frac{\tr\left(e^{-2\pi n H_{\rm an}}\right)}{\lt(\tr\left(e^{-2\pi H_{\rm an}}\right)\rt)^n}
\]
(this is to be compared with \eqref{expValOrastraceno2}). It is a simple matter to check that this gives the same results as those obtained here, including logarithmic renormalization and VEVs.
\item
The CFT renormalization formula \eqref{Zexpov}, or its equivalent for branch-point twist field (see \eqref{twistprod}), involves a logarithmic factor that was missed in the original work \cite{CW94_EE}. The reason for this omission is simple. By the mode decomposition, either in angular quantization or directly in the path-integral formalism, one obtains, for the real massless free boson, the expression
\[
	\langle \T^\ep(0)\tT^\ep(x)\rangle
	=\prod_{k=1}^\infty \frac{(1-q^k)^n}{1-q^{nk}}\qquad
	\mbox{(real massless free boson)}
\]
(see for instance \eqref{Zprod} and the discussion surrounding it), where the limit $\ep\to0$ corresponds to $q\to1$. The usual technique is to assume that we can convert the logarithm of this expression into an integral over $y=k\,|\log q|$. This gives the leading divergency times a factor reproducing the correct dimension. However, because the integrand is divergent at small $y$, the corrections to this leading divergency are {\em not} of order one, but rather involve $\log |\log q|$. The Mellin transform technique (see Appendix \ref{AAPPMel}) gives these corrections. This is different from the case of the Dirac fermion, where the integrand is not divergent at small $y$ and thus the naive replacement of a sum by an integral is correct.
\item The two-point function of the branch-point twist field was studied using a form factor expansion in \cite{BC16_BranchP}. It was found, by numerically analyzing the form factor series, that at short distances, the usual power law is corrected by a factor that tends to zero, which, it was argued, should be identified as a logarithmic factor. This is in agreement with our result \eqref{ttov}, where the power of the logarithm is negative. It was also argued, by comparing with the work \cite{CCT1_2009}, that in the short-distance behavior of the universal scaling function for the EE, the corresponding double-logarithm should be of the form $\log(\log(L/\xi))$, with coefficient 1. Our calculation instead gives a coefficient $1/2$, see \eqref{S1fb4} (note that the term $-\log\log\xi$ in \eqref{S1fb3} was absent in \cite{BC16_BranchP} because the logarithmic renormalization of the field $\T^\ep$ was not considered). However, we must emphasize that the exact power in \eqref{ttov}, and the exact coefficient of $\log(\log(L/\xi))$ in \eqref{S1fb4}, are based on certain assumptions as explained around \eqref{aofje}. In our view the question of the exact coefficient remains unsettled, although the evidences of the present work and that of \cite{BC16_BranchP} indicate that it should be positive.
\item We have not specified the exact procedure by which the EE should be evaluated, on the harmonic chain, in order to lead to the universal results of Section \ref{sectE}. This is a rather delicate matter. For instance, in the massless harmonic chain, the non-compactness of the target space renders any direct numerical calculation unfeasible. One must fix the gauge freedom afforded by constant shifts of the field, and this may break the tensor structure of the Hilbert space ${\mathcal H} = {\mathcal H}_A \otimes {\mathcal H}_B$. The choice of Dirichlet boundary conditions we have made corresponds to a particular choice of gauge fixing, which renders the calculation in the massless case well defined. However, different gauge fixing will lead to different renormalization, and thus, double-logarithm terms in the renormalization procedure, are of a lesser degree of universality. For instance, as remarked at the end of Section \ref{QuantAn}, we show in appendix \ref{appneu} that with Neumann boundary conditions there is no logarithmic renormalization (no double-logarithm term in \eqref{dblimlog}, \eqref{S1fb}, \eqref{S1fb2} and \eqref{S1fb3}). Nevertheless, the scaling function $F(mx) = m^{-2h_n}\langle \T(x)\tT(0)\rangle$ in \eqref{S1fb3} is expected to be universal and independent of such a choice (and its short-distance behavior displays a double-logarithm term as per \eqref{S1fb4}).
\item Logarithmic contributions to the EE were first proposed in \cite{BCDLF_nonU14}, under the assumption that there is a finite number of primary fields and that $L_0$ has a non-diagonalizable Jordan form. The proposition agrees with the general form \eqref{S1fb2}, except that the coefficient of $\log(\log(L))$ was identified as a nilpotency power, hence an integer number. The free massless boson is logarithmic, but has an infinite number of primary fields, hence does not fit into the general scheme of \cite{BCDLF_nonU14}. It would be interesting to clarify the full relation between the present result \eqref{S1fb2} and the techniques of \cite{BCDLF_nonU14}.
\item The $\grp{U}(1)$ twist fields in the complex 1+1-dimensional Klein Gordon model were first studied in \cite{SMJ_79IV}, where differential equations of Painlev\'e type were found to describe their two-point functions (as obtained from their form factor expansions). Our results for the universal scaling function $F(mx) = m^{-2h_n}\langle \T(x)\tT(0)\rangle$ therefore should apply to the corresponding Painlev\'e tau-functions, and in particular, the exact VEV \eqref{VEVboson} be related to a Painlev\'e connection problem. It would be interesting to analyze this.
\end{enumerate}

Finally, it would be interesting to use the methods developed here for cases of interacting integrable QFT, and in particular those whose CFT are Wess-Zumino-Witten where there are symmetry currents associated to non-abelian groups. We also note that it would be interesting to study the method we have proposed, of solving a differential equation with respect to a continuous group parameter, for other fundamental constants such that the coupling constant $C_{\T \T}^{\T^2}$ that appears in the operator algebra of branch-point twist fields, $\T(x) \T(0)\sim C_{\T \T}^{\T^2} \T^2(0)+ \ldots$; a simple calculation for the Dirac fermion shows that this gives rise to the known coupling constant in this case.

As this paper was in preparation, we were made aware of independent results \cite{CJS_2016} where the result \eqref{S1fb2}, with in particular the same term $-(1/2)\log(\log(L))$, was obtained using techniques of loop partition functions, in certain limits of CFT minimal models (known to be logarithmic).

\section*{Acknowledgements}

The authors wish to thank D.~Bianchini and O.~ A.~Castro-Alvaredo for discussions and for their role in the initial stage of this project. BD thanks H.~Saleur, P.~Calabrese and the group of Statistical Physics at the Scuola Internazionale Superiore di Studi Avanzati (SISSA), Trieste, Italy, for discussions. BD acknowledges support from SISSA where parts of this work were done. OBF is supported by ``Fonds de Recherche du Qu\'{e}bec -- Nature et Technologies". BD is partially supported by EPSRC grant EP/P006132/1.

\begin{appendix}

\section{Asymptotic expansions and the Mellin transform}  \label{AAPPMel}
\label{MTransfo}

In this appendix, we recall some properties of the Mellin transform.   The asymptotic expansion needed in the computation of one-point and two-point functions in Section \ref{QuantAn} is obtained via the Mellin transform.  We also present in this appendix the details of the calculation for the asymptotic value   \eqref{Zexp}.   
 There are many good references on the Mellin transform, see for instance \cite[App. B]{FS_AC09} or  \cite{F_MT95, DZagier06}, from which the following presentation is mostly  inspired.
  
Let $f(x)$ be a smooth function defined on the positive real axis ($x>0$), which decays fast enough at $x=0$ and $x=\infty$.  Then the function
\be\label{eqdefMtr1}
f^\star(s) = \int_0^\infty x^{s-1} f(x) \dd x
\ee
is well defined and analytic for complex values of $s$ in an appropriate domain, and is defined beyond this domain by analytic continuation. This function is called the Mellin transform of $f(x)$; we will use the symbol $\mathfrak M$ for the map $f(x) \mapsto f^\star(s)$. Let $f(x) \sim x^a, x\rightarrow 0$  and $f(x) \sim x^b, x\rightarrow \infty$. Then the fundamental strip where the integral in \eqref{eqdefMtr1} is well defined and $f^\star(s)$ is analytic is given by $-a < \mathrm{Re}(s)<-b$.  The inverse of the Mellin transform, to be denoted by $\mathfrak{M}^{-1}: f^\star(s)\mapsto f(x)$, is given by
\be
f(x) = \frac{1}{2\pi \ii} \int_{\beta-\ii\infty}^{\beta+\ii\infty} x^{-s} f^\star(s) \dd s
\ee
for $\beta$ in the fundamental strip (and $\ii=\sqrt{-1}$).

The key observation is that this $\mathfrak M$ transforms the asymptotic expansion of $f(x)$ around zero and around infinity to singularities of $f^\star(s)$. Conversely, $\mathfrak{M}^{-1}$ transforms singularities of $f^\star(s)$ (outside the fundamental strip) to terms in the asymptotic expansion of $f(x)$. The set of all singularities of $f^\star(s)$ thereby gives rise to the full asymptotic expansion of $f(x)$. 
There is a simple dictionary between the poles of $f^\star(s)$ and the terms contributing in the asymptotic expansion of $f(x)$.  This can be used as follows: under the inverse Mellin transform, we have the correspondance
\be\label{MT_dico1}
\frac{1}{(s-s_0)^{k+1}}  \quad \xmapsto{\mathfrak M^{-1}} \quad  \pm \,  \frac{(-1)^k}{k!} \, x^{-s_0} \, (\log x)^k
\ee
where the $+ (-)$ is for a pole on the left (right) of the fundamental strip.  

There are a number of properties of the Mellin transform which tell how $f^\star(s)$ changes when $f(x)$ is modified. We will mainly use the property that $\mathfrak M$ is a linear transformation and that $f(kx)\xmapsto{\mathfrak M} k^{-s}f^\star(s)$, in particular
\be\label{MT_eq_psumno1}
\sum_{k=1}^\infty f(k x)  \quad \xmapsto{\mathfrak M} \quad \zeta(s) f^\star(s)
\ee
where the Riemann zeta function $\zeta(s)=\sum_{k\in\ZZ_+} k^{-s}$ is involved.

We now consider a precise example of a computation of an asymptotic expansion using the Mellin transform,  which will connect with the expression \eqref{Zprod}.  
Consider the following function
\be
P(x;t) = -\sum_{k=1}^{\infty} \log(1-te^{-kx}) = \sum_{k=1}^\infty \frac{t^k}{k} \, \frac{e^{-kx}}{1-e^{-kx}}
\ee
from which we are interested in finding the asymptotic expansion when $x\rightarrow0$.  The parameter $t$ is unspecified for now.  Note that the relation with \eqref{Zprod} is given by
\be\label{AppMTrellogZP1}
	\log \langle \T_{-\alpha}^\epsilon(0)\T_\alpha^\epsilon(x)\rangle_{\rm CFT} = P\bigl( \tfrac{2\pi^2}{\Delta} ;\, e^{2\pi \ii \alpha} \bigr) +  P\bigl( \tfrac{2\pi^2}{\Delta} ;\, e^{-2\pi \ii \alpha} \bigr) - 2  P\bigl( \tfrac{2\pi^2}{\Delta} ;\, 1 \bigr).
\ee
Using $e^{-x}/(1-e^{-x})= \sum_{k=1}^\infty e^{-kx}$ along with \eqref{MT_eq_psumno1}, we have
\be
 \frac{e^{-x}}{1-e^{-x}}  \quad \xmapsto{\mathfrak M} \quad \Gamma(s) \zeta(s), 
 \ee
 where $\Gamma(s)$ is the Gamma function and  where the fundamental strip is given by $\mathrm{Re}(s)>1$. We then obtain
\be\label{MTgenFctno9}
P(x; t )  \quad \xmapsto{\mathfrak M} \quad  P^\star(s;t)= \sum_{k=1}^\infty \frac{t^k}{k} k^{-s} \Gamma(s) \zeta(s) = \mathrm{Li}_{s+1}(t) \Gamma(s) \zeta(s)
\ee
where the function $\mathrm{Li}_{s+1}(t)$ is the polylogarithm function.    

For $t=1$, the polylogarithm function reduces to $\zeta(s+1)$, so that $ P^\star(s;1)= \zeta(s+1)\Gamma(s) \zeta(s)$.  This function has poles at $s=1,0,-1,-2, \ldots$, but since (using dictionary \eqref{MT_dico1}) the poles at $s=-1,-2,\ldots$ will only contribute in the asymptotic expansion of $P(x,1)$ as $O(x^1), O(x^2), \ldots$ as $x\rightarrow 0$, we may ignore them.    
At $s=1$, the only pole comes from $\zeta(s)$ so we have
\be
P^\star(s;1) \sim \frac{\zeta(2)}{s-1}, \quad s\rightarrow 1 \qquad \xmapsto{\mathfrak M^{-1}} \qquad P(x;1) \sim (+) x^{-1} (\pi^2/6) + \ldots, \quad x\rightarrow 0
\ee
where we have chose the $(+)$ sign because to pole is at left of the fundamental strip.  
At $s=0$ we have a pole of order two coming from $\Gamma(s)$ and $\zeta(s+1)$, so that
\be
P^\star(s;1) \sim \frac{\zeta(0)}{s^2} + \frac{\zeta'(0)}{s}, \quad s\rightarrow 0 
\qquad \xmapsto{\mathfrak M^{-1}} \qquad P(x;1) \sim  \ldots + 
\frac12 \log(x) -\frac12 \log(2\pi) + \ldots, \quad x\rightarrow 0
\ee 
where $\zeta'(s)=\partial_s \zeta(s)$.  Adding all the contributions, we thus have
\be\label{eqlogp1}
 P(x;1)  \sim \frac{(\pi^2/6)}{x} + \frac12 \log(x) -\frac12 \log(2\pi)+  O(x), \qquad x\rightarrow 0.
\ee

For general $t\neq 1$, the right-hand side of \eqref{MTgenFctno9} only has poles at $s=1,0$.  For $s=1$, there is a pole of order one with residue $\mathrm{Li}_2(t)$, and $s=0$ is a pole of order one with residue $-\tfrac12 \mathrm{Li}_1(t)$.  
We thus have
\be
P(x;t)\sim \frac1x \mathrm{Li}_2(t) - \frac12 \mathrm{Li}_1(t) +  O(x), \qquad x\rightarrow 0 \quad (t \neq1).  
\ee
Observe that one can not simply gets \eqref{eqlogp1} by setting $t=1$ in this last result, pointing to to the non-commutativity of the limits $x\to0$ and $t\to1$.

We end this appendix with \eqref{AppMTrellogZP1}.  The asymptotic expansion of this expression is obtained using the previous two results,
\be\begin{split}
\log\langle \T_{-\alpha}^\epsilon(0)\T_\alpha^\epsilon(x)\rangle_{\rm CFT} & \sim 
\frac{\Delta}{2\pi^2} \bigl( \mathrm{Li}_2(e^{2\pi \ii \alpha}) + \mathrm{Li}_2(e^{-2\pi \ii \alpha}) - 
\frac{\Delta }{6} - \frac12 \bigl( \mathrm{Li}_1( e^{2\pi \ii \alpha}) + \mathrm{Li}_1(e^{-2\pi \ii \alpha}) \bigr) 
\\
& \quad
- \log\Bigl( \frac{2\pi^2}{\Delta}\Bigr) + \log(2\pi) + O\Bigl(\frac{1}{\Delta} \Bigr),
\end{split}
\ee
as $\Delta\rightarrow \infty$.    Then using the following identities
\be\begin{split}
\mathrm{Li}_1(e^{2\pi \ii \alpha}) + \mathrm{Li}_1(e^{-2\pi \ii \alpha})  &= -\log \bigl( (1-e^{2\pi \ii \alpha}) (1-e^{-2\pi \ii \alpha}) \bigr) = - \log(4 \sin^2(\pi \alpha))\\
\mathrm{Li}_2(e^{2\pi \ii \alpha}) + \mathrm{Li}_2(e^{-2\pi \ii \alpha}) &= -\frac{(2\pi \ii)^2}{2} \mathrm{B}_2(\alpha)= 2\pi^2 \bigl( \alpha^2-\alpha+\frac16 \bigr)
\end{split}
\ee
where $\mathrm{B}_2(\cdot)$ is the Bernoulli polynomial of order $2$, we finally obtain
\be\label{logZA}
\log\langle \T_{-\alpha}^\epsilon(0)\T_\alpha^\epsilon(x)\rangle_{\rm CFT}  \sim
-\alpha(1-\alpha)\Delta  + \log(\Delta) + \log\Bigl[\frac2\pi |\sin( \pi \alpha)|\Bigr]
+ O\Bigl(\frac{1}{\Delta} \Bigr).
\ee
Note that this asymptotic expansion gives a term of the form $\log(\Delta)$, and since $\Delta$ is itself proportional to $\log(\epsilon)$ we see that the two-point function  contains a term in $\log(\epsilon)$.

\section{The case of the Neumann boundary condition}\label{appneu}

We look at the case of free boundary condition for the complex boson in the angular quantization formalism.  First consider the two-point function in CFT (following the steps of paragraph \ref{s432}).   We wish to impose
\be\label{eqBC1}
\partial_\eta \phian(\eta=\log(\epsilon/x))=  \partial_\eta \phian(\eta=\log(x/\epsilon))=0.
\ee  
We can write the Fourier mode decomposition:
\be
\phian(\eta)= \sum_{k\in\mathbb Z_{\geq 0}} (c_k + d^\dagger_k) \cos(\pi k \hat \eta/\Delta)
\ee
and this automatically satisfies \eqref{eqBC1}. Note that we now have a zero mode.  The Hamiltonian and charge in angular quantization stay the same as before, and the set of energies is sill given by
\be
\nu_k= \pi k/\Delta, \qquad k \in \mathbb Z_{\geq 0}.
\ee
Since $\nu_0=0$ the contribution of the zero modes coming from the Hamiltonian vanishes and we  then have
\be\label{eq_trproddnu1234}\begin{aligned}
\langle \T_{-\alpha}^\epsilon(0)\T_\alpha^\epsilon(x)\rangle_{\rm CFT} &=  \frac{ \tr \lt[  \prod_{k=0}^\infty \exp \Bigl( -2 \pi  (\nu_k^{\phantom\dag} + \ii \alpha) c_k^\dag c_k^{\phantom\dag} - 2 \pi ( \nu_k^{\phantom\dag}  - \ii \alpha) d_k^\dag d_k^{\phantom\dag} \Bigr)\rt] }{ \tr\lt[  \prod_{k=0}^\infty \exp \Bigl( -2 \pi  \nu_k^{\phantom\dag}  c_k^\dag c_k^{\phantom\dag} - 2 \pi  \nu_k^{\phantom\dag}  d_k^\dag d_k^{\phantom\dag} \Bigr) \rt]}  \\
&=   \frac{ \tr_{\rm bos} \lt[ \exp \bigl( -2 \pi \ii \alpha \h n\bigr)\rt]  \tr_{\rm bos} \lt[ \exp \bigl( 2 \pi \ii \alpha \h n\bigr)\rt]}{\mathrm K_\infty}  \times
\nonumber\\ & \qquad \times
 \prod_{k=1}^{\infty} \frac{ \tr_{\rm bos} \lt[ \exp \Bigl( -2 \pi  (\nu_k + \ii \alpha) \h n\bigr)\rt]}{ \tr_{\rm bos} \lt[\exp \Bigl( -2 \pi  \nu_k  \h n\Bigr)\rt] } \,
 \frac{ \tr_{\rm bos} \lt[ \exp \Bigl( -2 \pi  (\nu_k - \ii \alpha) \h n\Bigr)\rt]}{ \tr_{\rm bos} \lt[\exp \Bigl( -2 \pi  \nu_k  \h n\Bigr) \rt]}.
\end{aligned}
\ee
Note that there is an infinite constant ${\mathrm K_\infty}$ due to the trace over the identity (which is infinite-dimensional) in the Fock space.  This therefore necessitates an extra regularization. We choose the regularization according to which it {\em diverges along with $|\log\ep|$} and set to:
\be
	{\mathrm K_\infty} = \frac{|\log\epsilon|}{(2\pi)^2}.
\ee

The product starting from $k=1$ to $k=\infty$ was already evaluated in the asymptotic limit (see paragraph \ref{s432} and Appendix \ref{AAPPMel}), and the remaining contribution is
\be
\tr_{\rm bos} \lt[ \exp \bigl( -2 \pi \ii \alpha \h n\bigr)\rt]  \tr_{\rm bos} \lt[ \exp \bigl( 2 \pi \ii \alpha \h n\bigr)\rt]   = \frac{1}{(1-t)(1-t^{-1})} = \frac{1}{4 \sin^2(\pi \alpha)}, \qquad t=e^{2\pi \ii \alpha}.
\ee
All together, this gives (as $\ep\to0$)
\be
\langle \T_{-\alpha}^\epsilon(0)\T_\alpha^\epsilon(x)\rangle_{\rm CFT}  \sim \frac{4\pi}{\sin(\pi \alpha)} \bigl( \frac{\epsilon}{x} \bigr)^{2h_\alpha}
\ee
from which we deduce the normalization (and regularization) constants
\be
\ell_2=0, \qquad c_\alpha=\sqrt{\frac{ \sin(\pi \alpha)}{4\pi}}.
\ee

Now we look at the one-point function in the massive theory (following the steps of paragraph \ref{s433}), using the same boundary condition:
\be
\partial_\eta \phian(\eta=\log \epsilon) = 0.
\ee
The important result to point out here is that this condition imposes on the set of energies the following condition:
\be
\nu_k = \frac{(2\ii)^{-1} \log \mathrm Z(\nu_k) - \pi k}{\log(m \epsilon/2)}, \qquad k=0,1,2,\ldots
\ee
with
\be
\mathrm Z(\nu) = \frac{\Gamma(\ii \nu)}{\Gamma(-\ii \nu)}.
\ee
At leading order in $u=-\pi/\log(m\epsilon/2)$ when $\nu$ is small we have
\be
\nu_k = \frac{(2\ii)^{-1} (- \ii \pi)- \pi k}{\log(m \epsilon/2)} =u \bigl( k+\frac12\bigr), \qquad k=\mathbb Z_{\geq 0}.
\ee
Thus, at leading order, we have to evaluated the asymptotic of
\be
A_1=\prod_{k=0}^\infty  \frac{(1-q^{k+1/2})^2}{(1-tq^{k+1/2})(1-t^{-1} q^{k+1/2}) }, \qquad q=e^{-2\pi u}
\ee
when $q\rightarrow1$.  This can be done again using the Mellin transform (Appendix \ref{AAPPMel}).  Set $q=e^{-x}$ and consider
\be
P(x;t) = - \sum_{k=0}^\infty  \log(1-t e^{-x(k+1/2)}), \qquad x\rightarrow 0
\ee
so that
\be
\log A_1=P(2\pi u; t)+ P(2\pi u;t^{-1}) - 2 P(2\pi u;1).
\ee
Using the same notation as in Appendix \ref{AAPPMel}, we have for the Mellin transform,
\be
P^\star(s;t)=\sum_{k=1}^\infty \frac1k t^k \zeta^{\mathrm H}(s;1/2) \Gamma(s) = \rm{Li}_{s+1}(t) \zeta^{\mathrm H}(s;1/2) \Gamma(s)
\ee
with (now) the Hurwitz zeta function
\be
 \zeta^{\mathrm H}(s;a)= \sum_{k\geq 0}\frac{1}{(k+a)^s}.
\ee
Using the identity
\be
 \zeta^{\mathrm H}(s;1/2) = (2^s-1) \zeta(s) ,
\ee
it is straightforward to evaluate to poles of $P^\star(s;t)$.  Recall that ${\rm Li}_{s+1}(1)=\zeta(s+1)$.  For $t=1$, only the contribution of the poles at $s=0,1$ will contribute to the asymptotic of $x\rightarrow 0$, and since
\be
P^\star(s;1)\sim  \frac1s \bigl( -\frac12 \log(2) \bigr), \quad s\rightarrow 0; \qquad P^\star(s;1)\sim\frac{\zeta(2) \Gamma(1)}{s-1}, \quad s\rightarrow 1
\ee
we have
\be
P(x;1)\sim  \frac{\pi^2/6}{x}- \frac12 \log(2) + O(x), \qquad x\rightarrow 0.
\ee
Likewise, from the same pole contributions at generic $t\neq1$,
\be
P(x;t)\sim  \frac{1}{x} {\rm Li}_{2}(t) + O(x), \qquad x\rightarrow 0.
\ee
The regularized one-point function of the branch-point twist field is thus
\be
\log \langle \T_\alpha^\epsilon \rangle = \alpha(1-\alpha) \log(m\epsilon/2) +  \log(2)  + O(1)
\ee
where the remaining $O(1)$ term is the integral in the exponential in \eqref{Zm}. Note that this can only be renormalized with $\ell_1=0$ (that is, there is no $\log(\log(\cdot))$ term).  With the constant $c_\alpha$ above, we finally have
\be
\langle \T_\alpha \rangle = m^{h_\alpha} 2^{-\alpha(1-\alpha)} \sqrt{\frac{\sin(\pi \alpha)}{\pi}}
\exp\lt[
	\int_0^\infty \frac{\dd t}{t} e^{-t}
	\lt( \frac{2\sinh \bigl( \frac{t\alpha} 2 \bigr) \,\sinh \bigl( \frac{t(1-\alpha)}2\bigr)}{(1-e^{-t})\, \sinh \bigl( \frac t2 \bigr)} -h_\alpha\rt)
	\rt].
\ee
This is a solution to the differential equation for the VEV, and exactly agrees with the VEV \eqref{VEVboson} obtained in the Dirichlet case.

\end{appendix}

\flushleft
 \bibliography{vev_twist-v2}
 \bibliographystyle{acm}

\end{document}